\documentclass{article}
\usepackage{graphicx,color,amsmath,fullpage}

\numberwithin{equation}{section}
\numberwithin{table}{section}

\begin{document}

\title{The critical state in thin superconductors as a mixed boundary value problem: analysis and solution by means of the Erd\'elyi-Kober operators}

\author{Roberto Brambilla$^1$, Francesco Grilli$^2$\\ \\
$^1$\small Ricerca sul Sistema Energetico Ð RSE S.p.A.,Via Rubattino 54, 20134 Milano, Italy \\
$^2$\small Karlsruhe Institute of Technology, Hermann-von-Helmholtz-Platz 1, 76344 Eggenstein-Leopoldshafen, Germany}
\date{}

\maketitle

\begin{abstract}
With this paper we provide an effective method to solve a large class of problems related to the electromagnetic behavior of thin superconductors. Here all the problems are  reduced to finding the weight functions for the Green integrals that represent the magnetic field components; these latter must satisfy the mixed boundary value conditions that naturally arise from the critical state assumptions. The use of the Erd\'elyi-Kober operators  and of the Hankel transforms (and mostly the employment of their composition properties) is the keystone to unify the method toward the solution. In fact, the procedure consists always of the same steps and does not require any peculiar invention. For this reason the method, here presented in detail for the simplest cases that can be handled in analytical way (two parts boundary), can be directly extended to many other more complex geometries (three or more parts), which usually will require a numerical treatment.
In this paper we use the operator technique to derive the current density and field distributions in perfectly conducting and superconducting thin discs and tapes subjected to a uniform magnetic field or carrying a transport current. Although analytical expressions for the field and current distributions have already been found by other authors in the past by using several other methods, their derivation is often cumbersome or missing key details, which makes it difficult for the reader to fully understand the derivation of the analytical formulas and, more importantly, to extend the same methods to solve similar new problems. On the contrary, the characterization of these cases as mixed boundary conditions has the advantage of referring to an immediate and na\"{i}ve translation of physics into a consistent mathematical formulation whose possible extension to other cases is self-evident.

[The final publication is available at http://link.springer.com/article/10.1007/s00033-011-0185-5.]
\end{abstract}



\section{Introduction}


Following Bean's original works on the magnetization of hard superconductors (see~\cite{J:1962:Bean62} and ~\cite{J:1964:Bean64}), the so-called critical state model has been the topic of an impressive amount of works aimed at finding the current density and magnetic field distributions and at computing the ac losses in superconductors.

In 1970 Norris proposed a method to compute the ac losses in elliptical and infinitely thin tapes carrying a transport current~\cite{J:1970:Norris70}. His formulas of the ac losses as a function of the current amplitude are still widely used to check experimental data and to find reasons of discrepancy between the losses of real samples and the theoretical predictions. 

In the same year Halse derived the formula for computing the ac losses in a thin superconducting strip subjected to an external magnetic field perpendicular to its face~\cite{J:1970:Halse70}.

More than twenty years later, with almost simultaneous works, Brandt and Indenbom~\cite{J:1993:Brandt93} and Zeldov {\it et al.}~\cite{J:1994:Zeldov94} modified Bean's approach and were able to find analytical expressions for the current density and the field profiles as well for the ac losses of infinitely thin tapes subjected to an external transport current, a uniform magnetic field, or a combination of the two. The expression for the losses as a function of the amplitude of the applied field derived in~\cite{J:1970:Halse70,J:1993:Brandt93,J:1994:Zeldov94} are currently widely used to evaluate the loss performance of rare earth-based coated conductors (the so called second generation of high-temperature superconductor tapes), where the superconductor can be often approximated as an infinitely thin strip.

Mainly motivated by SQUID research, Mikeenko {\it et al.}~\cite{J:1993:Mikheenko93} applied the critical state for determining the current and field distribution in thin superconducting discs subjected to external perpendicular fields. Their results were obtained as a superposition of the classical distributions obtained from a limit process of very flat ellipsoids. The same idea was also followed by Zhou {\it et al.}~\cite{J:1993:Zhu93}, who improved the calculation of the hysteretic loop. 

The simplest description of the superconductors is the critical state model (CSM), according to which the current density magnitude cannot exceed a critical value $J_c$ and only in the regions where this limit is reached there is penetration of the magnetic flux and loss dissipation. In the regions where the current density magnitude is lower than $J_c$, the magnetic field is null. In this description of the superconducting state the electric resistivity is not defined and the usual methods to solve Maxwell equations in differential form cannot be applied since the current density is not locally linked to the electric field. On the other hand, the current tends to assume a distribution such as to screen as much as possible the superconductor from the penetration of externally applied fields as well as of the self-field produced by the current itself. Given the global character of this request, the differential formulation of the problem is naturally substituted by an integral formulation, whose unknown function is the current density distribution, subjected to the constraints of the critical state. The Biot-Savart law adapted to the geometry of the considered problem has been therefore the unavoidable starting point of any developments on this topic. The (sometimes remarkable) differences between the various developed methods ultimately consist in the differences between the techniques to solve the singular integral equations arising from this approach.
Many of those methods are based, in the two-dimensional cases, on the search for the appropriate conformal maps that can reduce the investigated geometries to the reference ones that can be treated. Often these methods are quite involute and difficult to be generalized.

In order to overcome these difficulties, we propose a new method that substitutes the Biot-Savart starting point with the more natural characterization of the critical state as a mixed boundary problem for the magnetic field. Expressing its components as integrals of the  appropriate Green functions for  the geometry and symmetry of the conductors, the coupled integrals equations that naturally result by applying the mixed boundary conditions can be transformed in an equivalent operator system by the use of the Erd\'elyi-Kober (EK) operators, which have been devised and developed for solving systems of coupled integral equations in a standard way~\cite{J:1962:Sneddon62}. Before their invention in the forties and their sporadic usage until the seventies, multiple integral equations were solved by ingenious trial solutions based on extensive use of special functions and their integrals, as in the case of Bessel fuctions, the omnipresent Weber-Shafheitling or Sonine integrals. By the translation in the operator frame all these intricacies disappear since they are hidden in the operator inversion and combination properties. The reader can find a very thorough description of the EK operators in the dedicated monography by Sneddon~\cite{J:1962:Sneddon62} and in the excellent article by Cooke~\cite{J:1963:Cooke63}, which considerably extends Sneddon's results.

Since to the best of the authors' knowledge this method has never been used in the field of applied superconductivity, in this work we present the application of the method to solve old problems, whose solution has often resulted not satisfying and/or passed from an article to another without thorough verification. 

Since the operator method requires an integral formulation of the problem, the starting point of this method consists in the search for an integral expression of the magnetic field that uses Biot-Savart law and the symmetries generating the mixed boundary conditions. When the real axis $(0,\infty)$ acts as boundary and the boundary conditions split it into two parts, each of them characterized by an integral equation, one will deal with a dual system. When the real axis $(0,\infty)$ is divided into three or more parts, one will deal with a triple or $n$-ple system.

\section{Perfectly conducting disc in uniform magnetic field}
\begin{figure}
\centering
\includegraphics[width=0.45\textwidth] {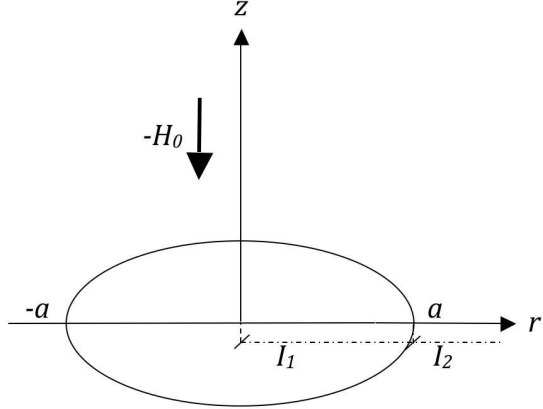}
\caption{Perfectly conducting disc in uniform magnetic field}
\label{fig:disc_pc}
\end{figure}

In order to show how the operator method works, we begin our report considering the very simple case of determining the currents induced in a perfectly conducting disc ($\sigma=\infty$) of radius $a$ subjected to a uniform magnetic field $-H_0$ perpendicular to the disc plane -- see figure~\ref{fig:disc_pc}. When the field is applied, circular eddy currents rise in the disc to oppose the inducing field. Since the disc is a perfect conductor, these currents completely null the resulting transverse magnetic field in the whole disc. 
According to the Green formulas (\ref{eq:appB_circ1}-\ref{eq:appB_circ2}) of appendix B, the azimuthal induced sheet current (this latter defined as the current density integrated over the thickness) will be 
\begin{align}\label{eq:Jphidiscp0}
J_\phi  (r) = 2H_r(r) \quad \quad (0<r<a) 
\end{align}
where $H_r(r)$ has to be determined from the following mixed boundary condition for the magnetic field

\begin{align}
\label{eq:Hzr}
H_z (r)&= S_{0,0} \psi (r) = H_0   \quad \quad(0 < r < a) \\
\label{eq:Hrr}
H_r (r) &= S_{1/2,0} \psi (r) = 0  \quad \quad (a < r < \infty).
\end{align}

The first equation expresses the reaction field, the latter describes the absence of the radial component of the field outside the disc, due to the symmetry of the problem. Consequently the radial axis is naturally divided in the two intervals $I_1 :(0,a) \cup I_2 :(a,\infty )$, and the unknown function $\psi(r)$ satisfies two different equations in each of them.
We shall call equation~(\ref{eq:Hzr}) and (\ref{eq:Hrr}) a dual operator system.
Points like $a$ where the boundary conditions change are called singular boundary points, since in their vicinity the solution often  presents strong discontinuities. Depending on their physical significance these discontinuities can be unacceptable, especially if they are in the interior of the conductors. If this is the case, the singular points have to be displaced along the axis until the unsound behavior disappears. Since the inversion of an equation like $S_{\eta ,\alpha } f = g$ requires the knowledge of the right term $g$ on the whole domain $(0 < r < \infty ) = I_1  \cup I_2 $, none of the two equations can be immediately inverted. One has to transform them, applying the EK operators, so that the resulting equivalent equations have the same left term. In this way the system is reduced to a unique operator equation over the whole axis and hence can be inverted. This is the leading idea behind the operator method.
In the next section we shall explain in detail how this task can be accomplished.

\section{Dual operator system, three mode reduction}
\label{sec:dual_operator}
Let us consider the following general Hankel operator system
\begin{align}\tag{i}
& S_{E,A} \psi  = f \label{eq:Sea1}\\
\tag{ii}
& S_{\overline{E},\overline{A}} \psi  = g \label{eq:Sea2}
\end{align}
in the case of the simplest fractioned boundary value problem, when the positive real axis is divided in two intervals by a point $a$, i.e. $I_1:(0,a) \cup I_2:(a,\infty)$. We can set $f$ as the sum ($\dot+$) of two disjointed parts, each of them  being defined only in the interval of its own index,  $f=f_1\dot+f_2$. If we know the couple of parts $(f_1,f_2)$, as in non-mixed problems, the solution $\psi$ is obtained from~(\ref{eq:Sea1}) by direct inversion by applying~(\ref{eq:Setaalpha_appA})
\begin{equation*}
\psi  = S_{E + A, - A} f = \left( {\begin{array}{*{20}c}
   a  \\
   0  \\
\end{array}} \right)S_{E + A, - A} f_1  + \left( {\begin{array}{*{20}c}
   \infty   \\
   a  \\
\end{array}} \right)S_{E + A, - A} f_2  = \psi _1  + \psi _2 
\end{equation*}
and the second equation is redundant. It is worth noting that each of the two terms $\psi_i$ is defined on the whole positive real axis, not just on one single interval.
Similarly, if $g=g_1\dot+g_2$ and the couple $(g_1,g_2)$ is known, from~(\ref{eq:Sea2}) we can immediately write
\begin{equation*}
\psi  = S_{\overline{E} + \overline{A}, - \overline{A}} g = \left( {\begin{array}{*{20}c}
   a  \\
   0  \\
\end{array}} \right)S_{\overline{E} + \overline{A}, - \overline{A}} g_1  + \left( {\begin{array}{*{20}c}
   \infty   \\
   a  \\
\end{array}} \right)S_{\overline{E} + \overline{A}, - \overline{A}} g_2  = \overline{\psi}_1  + \overline{\psi}_2. 
\end{equation*}
The direct inversion of~(\ref{eq:Sea1}) or (\ref{eq:Sea2}) is generally impossible when, as in the case of mixed problems, one knows parts of different functions in different intervals, as for example the couple $(f_1,g_2)$. Obviously, if the $S$ operators are identical ($E=\overline{E}$, $A=\overline{A}$), it is possible to derive
\begin{equation*}
\psi = \left( {\begin{array}{*{20}c}
   a  \\
   0  \\
\end{array}} \right)S_{E + A, - A} f_1  + \left( {\begin{array}{*{20}c}
   \infty   \\
   a  \\
\end{array}} \right)S_{E + A, - A} g_2 
\end{equation*}
or, if the couple $(f_2,g_1)$ is known,
\begin{equation*}
\psi  = \left( {\begin{array}{*{20}c}
   a  \\
   0  \\
\end{array}} \right)S_{E + A, - A} g_1  + \left( {\begin{array}{*{20}c}
   \infty   \\
   a  \\
\end{array}} \right)S_{E + A, - A} f_2.
\end{equation*}
When on the contrary the operators have different indices (as in the case of the disc), in order to carry out the inversion, one has to transform the dual system~(\ref{eq:Sea1}-\ref{eq:Sea2}) into an equivalent one with the same first members. By using the combination properties of the EK operators (see appendix~\ref{subsec:properties}), this task can be done in more than a unique way, as detailed below.
\subsection*{Mode I}

We rewrite the system changing index notation
\begin{align*}
& S_{e+a,b} \psi = f  & (E=e+a, A=b) \\
& S_{\overline{e},\overline{a}} \psi  = g & (\overline{E}=\overline{e}, \overline{A}=\overline{a})
\end{align*}
Let us multiply the first equation by $K_{e,a}$ and the second one by $I_{\overline{e} + \overline{a},\overline{b}}$, respectively
\begin{align*}
& K_{e,a} S_{e + a,b} \psi   =  S_{e,a + b} \psi = K_{e,a} f
\\
& I_{\overline{e} + \overline{a},\overline{b}} S_{\overline{e},\overline{a}} \psi  =  S_{\overline{e},\overline{a} + \overline{b}} \psi = I_{\overline{e} + \overline{a},\overline{b}} g.
\end{align*}
Requiring the identity of the middle members we shall have $e=\overline{e}$ and $a+b=\overline{a}+\overline{b}$, so that coming back to the original indexes we obtain
\begin{align*}
& a = E - \overline{E},\;e = \overline{E},\;b = A
\\
& \overline{a} = \overline{A},\;\overline{e} = \overline{E},\;\overline{b} = E - \overline{E} + A - \overline{A}
\end{align*}
and the equivalent system with the same first member will be
\begin{equation}\label{eq:system_becomes1}
S_{\overline{E},E - \overline{E} + A} \psi  = \varphi  = \left\{ {\begin{array}{*{20}c}
   {K_{\overline{E},E - \overline{E}} f}  \\
   {I_{\overline{E} + \overline{A},(E + A) - (\overline{E} + \overline{A})} g}  \\
\end{array}} \right.
\end{equation}
which, once inverted, is
\begin{align}
\nonumber
& f = K_{E,\overline{E} - E} \varphi  \\
\nonumber
& g = I_{E + A,(\overline{E} + \overline{A}) - (E + A)} \varphi  \\
& \psi  = S_{E + A,\;\overline{E} - (E + A)} \varphi  \label{eq:psi_method1}.
\end{align}
Developing it on the two intervals and considering the partitioned functions we can write (we omit the indices for sake of clarity)
\begin{align}
\nonumber
f_1 (r) &= \left( {\begin{array}{*{20}c}
   a  \\
   r  \\
\end{array}} \right)K\varphi _1 (r) + \left( {\begin{array}{*{20}c}
   \infty   \\
   a  \\
\end{array}} \right)K\varphi _2 (r)
& (r \in I_1) 
\\
f_2 (r) &= \left( {\begin{array}{*{20}c}
   \infty   \\
   r  \\
\end{array}} \right)K\varphi _2 (r)
& (r \in I_2) \label{eq:f2_method1}
\\
g_1 (r) &= \left( {\begin{array}{*{20}c}
   r  \\
   0  \\
\end{array}} \right)I\varphi _1 (r)
& (r \in I_1)  \label{eq:g1_method1}
\\
\nonumber
g_2 (r) &= \left( {\begin{array}{*{20}c}
   a  \\
   0  \\
\end{array}} \right)I\varphi _1 (r) + \left( {\begin{array}{*{20}c}
   r  \\
   a  \\
\end{array}} \right)I\varphi _2 (r)
& (r \in I_2).  
\end{align}
From this system we can try to recover the two parts of $\varphi$ from the knowledge of a couple of functions of the first members.
In case the couple $(f_2,g_1)$ is known, we can directly invert equations~(\ref{eq:f2_method1}) and~(\ref{eq:g1_method1}) using~(\ref{eq:inv_xa}) and (\ref{eq:inv_bx}), since the operators have a variable limit
\begin{align*}
\varphi _1 (r) &= \left( {\begin{array}{*{20}c}
   r  \\
   0  \\
\end{array}} \right)I^{ - 1} g_1 (r) & (r \in I_1)
\\
\varphi _2 (r) &= \left( {\begin{array}{*{20}c}
   \infty   \\
   r  \\
\end{array}} \right)K^{ - 1} f_2 (r) & (r \in I_2)
\end{align*}
and, according to~(\ref{eq:psi_method1}), we obtain the sought solution
\begin{equation}
\nonumber
\psi (r) = \left( {\begin{array}{*{20}c}
   a  \\
   0  \\
\end{array}} \right)S\varphi _1 (r) + \left( {\begin{array}{*{20}c}
   \infty   \\
   a  \\
\end{array}} \right)S_{} \varphi _2 (r).
\end{equation}
Even though this mode is particularly convenient when the couple $(f_2,g_1)$ is known, it can also be used when the couple $(f_1,g_2)$ is known; in that case it will generate a Fredholm integral equation of second kind, whose analytical solution may be a difficult exercise.

\subsection*{Mode II}

We rewrite the initial system changing index notation as follows
\begin{align}\tag{i}
& S_{e,a} \psi = f  & (E=e, A=a) \label{eq:Sea1ch}\\
\tag{ii}
& S_{\overline{e}+\overline{a},\overline{b}} \psi  = g & (\overline{E}=\overline{e}+\overline{a}, \overline{A}=\overline{b}) \label{eq:Sea2ch}
\end{align}
Let us multiply ~(\ref{eq:Sea1})  by $I_{e + a,b}$ and~(\ref{eq:Sea2}) by $K_{\overline{e},\overline{a}}$, respectively
\begin{align*}
I_{e + a,b} S_{e,a} \psi   =  S_{e,a + b} \psi  = I_{e + a,b} f
\\
K_{\overline{e},\overline{a}} S_{\overline{e} + \overline{a}} \psi  =  S_{\overline{e},\overline{a} + \overline{b}} \psi  = K_{\overline{e},\overline{a}} g.
\end{align*}
Requiring the identity of the middle members we must have $e=\overline{e}$ and $a+b=\overline{a}+\overline{b}$, so that coming back to the original indexes we obtain
\begin{align*}
a = A,\;e = E,\;b = \overline{E} - E + \overline{A} - A
\\
\overline{a} = \overline{E} - E,\;\overline{e} = E,\;\overline{b} = \overline{A}
\end{align*}
and the system becomes
\begin{equation}\label{eq:system_becomes2}
S_{E, \overline{E}- E + \overline{A}} \psi  = \varphi  = \left\{ {\begin{array}{*{20}c}
   {I_{E + A,(\overline{E} + \overline{A}) - (E + A)} f}  \\
   {K_{E, \overline{E} - E} g}  \\
\end{array}} \right.
\end{equation}
which by formal inversion gives
\begin{align}
\nonumber
f &= I_{\overline{E} + \overline{A},\;(E + A) - (\overline{E} + \overline{A})} \varphi \\
\nonumber
g &= K_{\overline{E},\;E - \overline{E}} \varphi 
\\
\psi  &= S_{\overline{E} + \overline{A},\;E - (\overline{E} + \overline{A})} \varphi  \label{eq:psi_method2}.
\end{align}
Writing the system explicitly on the two intervals, we shall have (we omit the indices for sake of clarity)
\begin{align}
f_1 (r) &= \left( {\begin{array}{*{20}c}
   r  \\
   0  \\
\end{array}} \right)I\varphi _1 (r)
& (r \in I_1) \label{eq:f1_method2}
\\
\nonumber
f_2 (r) &= \left( {\begin{array}{*{20}c}
   a  \\
   0  \\
\end{array}} \right)I\varphi _1 (r) + \left( {\begin{array}{*{20}c}
   r  \\
   a  \\
\end{array}} \right)I\varphi _2 (r)
& (r \in I_2) 
\\
\nonumber
g_1 (r) &= \left( {\begin{array}{*{20}c}
   a  \\
   r  \\
\end{array}} \right)K\varphi _1 (r) + \left( {\begin{array}{*{20}c}
   \infty   \\
   a  \\
\end{array}} \right)K\varphi _2 (r)
& (r \in I_1) 
\\
g_2 (r) &= \left( {\begin{array}{*{20}c}
   \infty   \\
   r  \\
\end{array}} \right)K \varphi _2 (r)
& (r \in I_2). \label{eq:g2_method2}
\end{align}
In the case the couple $(f_1,g_2)$ is known, equations~(\ref{eq:f1_method2}) and~(\ref{eq:g2_method2}) can be directly inverted
\begin{align}
\nonumber
\varphi _1 (r) &= \left( {\begin{array}{*{20}c}
   r  \\
   0  \\
\end{array}} \right)I^{ - 1} f_1 (r)
\\
\nonumber
\varphi _2 (r) &= \left( {\begin{array}{*{20}c}
   \infty   \\
   r  \\
\end{array}} \right)K^{ - 1} g_2 (r)
\end{align}
and from~(\ref{eq:psi_method2}) we obtain the sought solution
\begin{equation}
\nonumber
\psi (r) = \left( {\begin{array}{*{20}c}
   a  \\
   0  \\
\end{array}} \right)S\varphi _1 (r) + \left( {\begin{array}{*{20}c}
   \infty   \\
   a  \\
\end{array}} \right)S\varphi _2 (r).
\end{equation}
\subsection*{Mode III}
By using the previous two modes we can deduce also a third mode which involves the second members only, i.e a mode that can be used to directly find the missing part of the second members $f$ and $g$. Equating the second members of (\ref{eq:system_becomes1}) we obtain the following equation

\begin{equation}\label{eq:KEEE}
K_{\overline{E},E - \overline{E}} f = I_{\overline{E} + \overline{A},(E + A) - (\overline{E} + \overline{A})} g
\end{equation}
and equating the second members of~(\ref{eq:system_becomes2}) we obtain
\begin{equation}\label{eq:IEA}
I_{E + A,(\overline{E} + \overline{A}) - (E + A)} f = K_{E,\overline{E} - E} g.
\end{equation}
Developing them on the two intervals we obtain a system of equations from which we can try to recover the missing parts from the given ones without the need of the auxiliary function $\varphi$. If the knowledge of $f$ is complete, $\psi$ is derived from (\ref{eq:Sea1}); alternatively, if the knowledge of $g$ is complete, $\psi$ is obtained from (\ref{eq:Sea2}). In addition, if the parts of $f$ and $g$ represent physical quantities (as in our cases, the magnetic field components), equation (\ref{eq:KEEE}) and (\ref{eq:IEA}) directly express the link between them; if their knowledge is the goal of the problem, the computation of $\psi$ is no longer necessary.
Unfortunately this third mode generally suffers from the impossibility of applying the powerful double operators described in appendix~\ref{subsec:double_operators}.

\section{Perfectly conducting disc in uniform magnetic field -- solution}
We return to the dual system (\ref{eq:Hzr})-(\ref{eq:Hrr}) for the perfect conducting disc previously found and we rewrite it with the second members partitioned on the two intervals $I_1$ and $I_2$
\begin{align}
\label{eq:S00psi}\tag{i}
S_{0,0} \psi &= f_1 \dot+ f_2
\\
\label{eq:S120psi}\tag{ii}
S_{1/2,0} \psi &= g_1\dot+ g_2
\end{align}
where $f_1=H_0$, $g_2=0$, and $f_2$ and $g_1$ are unknown. According to the ideas of the preceding section, this system is equivalent to one of the three systems shown in table~\ref{tab:disc_unif} of appendix B, which can be solved by operator inversion.
\subsection{Second mode}
We start privileging the second mode, since we know the couple $(f_1,g_2)$. Expanding the  equations on the two intervals, we obtain the following four equations
\begin{align}
\label{eq:f1_disc}
H_0 &= \left( {\begin{array}{*{20}c}
   r  \\
   0  \\
\end{array}} \right)I_{1/2, - 1/2} \varphi _1 (r) & \quad \quad (r \in I_1 ) \\
\label{eq:f2_disc}
f_2 (r) &= \left( {\begin{array}{*{20}c}
   a  \\
   0  \\
\end{array}} \right)I_{1/2, - 1/2} \varphi _1 (r) + \left( {\begin{array}{*{20}c}
   r  \\
   a  \\
\end{array}} \right)I_{1/2, - 1/2} \varphi _2 (r) & \quad \quad (r \in I_2 )\\
\label{eq:g1_disc}
g_1 (r) &= \left( {\begin{array}{*{20}c}
   a  \\
   r  \\
\end{array}} \right)K_{1/2, - 1/2} \varphi _1 (r) + \left( {\begin{array}{*{20}c}
   \infty   \\
   a  \\
\end{array}} \right)K_{1/2, - 1/2} \varphi _2 (r) & \quad \quad (r \in I_1 )\\
\label{eq:g2_disc}
0 &= \left( {\begin{array}{*{20}c}
   \infty   \\
   r  \\
\end{array}} \right)K_{1/2, - 1/2} \varphi _2 (r) & \quad \quad (r \in I_2 ).
\end{align}
(The explicit expression of the EK operators presented here and in all the following paragraphs are listed in tables~\ref{tab:A1}-\ref{tab:A2} in appendix A.) From~(\ref{eq:g2_disc}) we evidently derive $\varphi _2  = 0 $, and from~(\ref{eq:f1_disc})
\begin{align}
\nonumber
\varphi _1 (r) = \left( {\begin{array}{*{20}c}
   r  \\
   0  \\
\end{array}} \right)I_{0,1/2} H_0  = \frac{2}{{\sqrt \pi  }}\frac{1}{r}\int\limits_0^r {\frac{{t^{} }}{{\sqrt {r^2  - t^2 } }}H_0 dt = } \frac{2}{{\sqrt \pi  }}H_0^{}.
\end{align}
Therefore the current density given by equation~(\ref{eq:Jphidiscp0}) becomes
\begin{align}
\nonumber
J_\phi  (r) &= 2g_1 (r) = 2\left( {\begin{array}{*{20}c}
   a  \\
   r  \\
\end{array}} \right)K_{1/2, - 1/2} \varphi _1 (r) \\
\label{eq:disc_Jz}
&= 
2\frac{{ - 1}}{{\sqrt \pi  }}\frac{d}{{dr}}\int\limits_r^a {\frac{t}{{\sqrt {t^2  - r^2 } }}\left( {\frac{2}{{\sqrt \pi  }}H_0 } \right)} dt = \frac{4}{\pi }H_0 \frac{r}{{\sqrt {a^2  - r^2 } }} \quad \quad (r \in I_1 ),
\end{align}
a well known result (see for example~\cite{J:1999:Shantsev99}) which we obtained here by means of two elementary integrals. Even though this result can also be obtained by more common methods, we present it as a first application to show the main distinctive steps necessary to utilize the operator method. Thanks to
its generality, this method can be easily extended to numerous other cases without any peculiar ingenuity. Essentially, from the above solution we can extract the following general procedure:
\begin{enumerate}
\item formulate the problem as a Hankel operator system;
\item choose an equivalent invertible system (same first members);
\item split the system on the intervals and solve using EK-operators properties.
\end{enumerate}
Let us now calculate the total magnetic field $H_t$ outside the disc, in the plane $z=0$. In order to do that, we add the applied field $-H_0$ to the reaction field given by equation~(\ref{eq:f2_disc})
\begin{align}
\nonumber
H_t (r) &= - H_0  + f_2(r) 
\\
\nonumber
&= - H_0  + \left( {\begin{array}{*{20}c}
   a  \\
   0  \\
\end{array}} \right)I_{1/2, - 1/2} \varphi _1 (r)
\\
\nonumber
& = - H_0  + \frac{1}{{\sqrt \pi  }}\frac{1}{r}\frac{d}{{dr}}\int\limits_0^a {\frac{{t^2 }}{{\sqrt {r^2  - t^2 } }}\left( {\frac{2}{{\sqrt \pi  }}H_0 } \right)} dt \\
& =  - H_0  + \frac{2}{\pi }H_0 \left( {\sin ^{ - 1} \frac{a}{r} - \frac{a}{{\sqrt {r^2  - a^2 } }}} \right)
\quad \quad (r \in I_2).
\end{align}
As it can be seen in figures~\ref{fig:J_disc} and \ref{fig:H_disc} (dashed curves), at the critical point $a$ (at the boundary of the disc) the sheet current $J_\phi(r)$ and the magnetic field $H_z(r)$ have a strong discontinuity and become infinite.  Nevertheless we can accept them as correct solutions: (i) from the mathematical point of view, because their spatial dependence keeps the integral of the sheet current and of the magnetic energy density (in any arbitrary region containing the edge of the disc) at a finite value~\cite{B:1991:VanBladel91}; (ii) from the physical point of view, because   these infinities result from the simplifying  assumption of zero thickness of the disc. If one considered a finite thickness, such singularities would disappear.
It is important to remark that in the present case the critical point is at the boundary of the conductor and the discontinuity can be accepted for the reasons mentioned above. On the other hand, when the critical point falls in interior points  (see section~\ref{sec:tape_pc_curr} and following ones) these infinities are no longer acceptable since the magnetic energy density integrals diverge. Their removal is the most difficult task in the case of numerical solutions. So, we have to add a fourth step to our procedure described above:
\begin{enumerate}
\setcounter{enumi}{3}
\item analyze results at critical points.
\end{enumerate}
Two examples of this analysis of the critical points are given in sections~\ref{sec:sc_disc_field} and \ref{sec:tape_sc_curr}, respectively.

\subsection{Third mode}
The same result can be obtained even more rapidly by employing the third mode. By developing the second equation of the third mode, we have the system
\begin{align}\label{eq:r0discp}
&\left( {\begin{array}{*{20}c}
   r  \\
   0  \\
\end{array}} \right)I_{0,1/2} (H_0 ) = \left( {\begin{array}{*{20}c}
   a  \\
   r  \\
\end{array}} \right)K_{0,1/2} g_1 
& (r \in I_1)
\\
\label{eq:a0discp}
&\left( {\begin{array}{*{20}c}
   a  \\
   0  \\
\end{array}} \right)I_{0,1/2} (H_0 ) + \left( {\begin{array}{*{20}c}
   r  \\
   a  \\
\end{array}} \right)I_{0,1/2} f_2  = 0
& (r \in I_2).
\end{align}
From~(\ref{eq:r0discp}) we get
\begin{equation}
\nonumber
g_1  = \left( {\begin{array}{*{20}c}
   a  \\
   r  \\
\end{array}} \right)K_{1/2, - 1/2} \left( {\begin{array}{*{20}c}
   r  \\
   0  \\
\end{array}} \right)I_{0,1/2} (H_0 ) = \left( {\begin{array}{*{20}c}
   a  \\
   r  \\
\end{array}} \right)K_{1/2, - 1/2} \left( {\frac{2}{{\sqrt \pi  }}H_0 } \right) = \frac{2}{\pi }H_0 \frac{r}{{\sqrt {a^2  - r^2 } }},
\end{equation}
and from~(\ref{eq:a0discp}) we get
\begin{equation*}
f_2  =  - \left( {\begin{array}{*{20}c}
   r  \\
   a  \\
\end{array}} \right)I_{0,1/2}^{ - 1} \left( {\begin{array}{*{20}c}
   a  \\
   0  \\
\end{array}} \right)I_{0,1/2} (H_0 ) =  - \left( {\begin{array}{*{20}c}
   r & a  \\
   a & 0  \\
\end{array}} \right)L_{0,1/2} \left( {H_0 } \right),
\end{equation*}
where we have introduced the double operator $L_{\eta ,\alpha }$ defined in appendix A, equation~(\ref{eq:double_Ina}). In the present case we get
\begin{equation*}
f_2 (r) =  - \frac{2}{\pi }\frac{{H_0 }}{{\sqrt {r^2  - a^2 } }}\int\limits_0^a {\frac{{\sqrt {a^2  - t^2 } }}{{r^2  - t^2 }}tdt} = H_0 \frac{2}{\pi }\left( {\sin ^{ - 1} \frac{a}{r} - \frac{a}{{\sqrt {r^2  - a^2 } }}} \right).
\end{equation*}
These results obtained with the third mode coincide with those obtained with the second mode. Had we developed the first equation of the third mode we would have obtained a Fredholm integral equation for $g_1$ whose kernel is a combination of EK operators that cannot be directly simplified using their properties
\begin{align*}
g_1 (r) &= \left\{ {\left( {\begin{array}{*{20}c}
   r  \\
   0  \\
\end{array}} \right)I_{1/2, - 1/2}^{ - 1} \left( {\begin{array}{*{20}c}
   \infty   \\
   a  \\
\end{array}} \right)K_{1/2, - 1/2}^{} \left( {\begin{array}{*{20}c}
   \infty   \\
   r  \\
\end{array}} \right)K_{1/2, - 1/2}^{ - 1} \left( {\begin{array}{*{20}c}
   a  \\
   0  \\
\end{array}} \right)I_{1/2, - 1/2}^{} } \right\}g_1 (r) \\
&+
 \left\{ {\left( {\begin{array}{*{20}c}
   r  \\
   0  \\
\end{array}} \right)I_{1/2, - 1/2}^{ - 1} \left( {\begin{array}{*{20}c}
   \infty   \\
   a  \\
\end{array}} \right)K_{1/2, - 1/2}^{} } \right\}H_0.
\end{align*}

\subsection{First mode}
We now try to apply the first mode to solve the problem. Using the equations of the first mode given in table~\ref{tab:disc} and developing the first on $I_1$ and the second on $I_2$, respectively, we have the system
\begin{align}
\label{eq:H0_discp}
H_0  &= \left( {\begin{array}{*{20}c}
   a  \\
   r  \\
\end{array}} \right)K_{0,1/2} \varphi _1  + \left( {\begin{array}{*{20}c}
   \infty   \\
   a  \\
\end{array}} \right)K_{0,1/2} \varphi _2 
& (r \in I_1) \\
\label{eq:0_discp}
0 &= \left( {\begin{array}{*{20}c}
   a  \\
   0  \\
\end{array}} \right)I_{0,1/2} \varphi _1  + \left( {\begin{array}{*{20}c}
   r  \\
   a  \\
\end{array}} \right)I_{0,1/2} \varphi _2
& (r \in I_2).
\end{align}
In equation (\ref{eq:0_discp}) we can invert only the second operator, so that, making again use of the double operators $L_{\eta ,\alpha }$, we obtain 
\begin{equation*}\label{eq:phi2discp}
\varphi _2  =  - \left( {\begin{array}{*{20}c}
   r  \\
   a  \\
\end{array}} \right)I_{0,1/2}^{ - 1} \left( {\begin{array}{*{20}c}
   a  \\
   0  \\
\end{array}} \right)I_{0,1/2} \varphi _1  =  - \left( {\begin{array}{*{20}c}
   r & a  \\
   a & 0  \\
\end{array}} \right)L_{0,1/2} \varphi _1.
\end{equation*}
On the other hand, from equation~(\ref{eq:H0_discp}) we can invert only the first operator
\begin{align*}\label{eq:phi1discp}
\varphi _1  &= \left( {\begin{array}{*{20}c}
   a  \\
   r  \\
\end{array}} \right)K_{0,1/2}^{ - 1} H_0  - \left( {\begin{array}{*{20}c}
   a  \\
   r  \\
\end{array}} \right)K_{0,1/2} \left( {\begin{array}{*{20}c}
   \infty   \\
   a  \\
\end{array}} \right)K_{0,1/2} \varphi _2  \\
&= \left( {\begin{array}{*{20}c}
   a  \\
   r  \\
\end{array}} \right)K_{1/2, - 1/2}^{} H_0  - \left( {\begin{array}{*{20}c}
   a & \infty   \\
   r & a  \\
\end{array}} \right)M_{0,1/2} \varphi _2.
\end{align*}
Eliminating $\varphi _2$,  we obtain
\begin{equation*}
\nonumber
\varphi _1  = \left( {\begin{array}{*{20}c}
   a  \\
   r  \\
\end{array}} \right)K_{1/2, - 1/2}^{} H_0  - \left( {\begin{array}{*{20}c}
   a & \infty   \\
   r & a  \\
\end{array}} \right)M_{0,1/2} \left( {\begin{array}{*{20}c}
   r & a  \\
   a & 0  \\
\end{array}} \right)L_{0,1/2} \varphi _1.
\end{equation*}%
By using the analytical expression of the operators, after changing the integration order, it is immediate to get the integral equation
\begin{equation*}
\varphi _1 = \frac{{H_0 }}{{\sqrt \pi  }}\frac{r}{{\sqrt {a^2  - r^2 } }} + \int\limits_0^a {K(r,t)\frac{{\sqrt {a^2  - t^2 } }}{{\sqrt {a^2  - r^2 } }}} \varphi _1 (t)dt,
\end{equation*}
where the kernel is 
\begin{equation}\label{eq:kernel_discp}
K(r,t) = \frac{2}{{\pi ^2 }}\frac{1}{{r^2  - t^2 }}\left( {r\ln \frac{{a - t}}{{a + t}} - t\ln \frac{{a - r}}{{a + r}}} \right).
\end{equation}
Multiplying both equation members by $(\sqrt \pi  /H_0 )\sqrt {a^2  - r^2 }$ and defining $\chi (r) = (\sqrt \pi  /H_0 )\sqrt {a^2  - r^2 } \varphi _1 (r)$, we finally obtain the Fredholm integral equation of second kind
\begin{equation}\label{eq:FredholmIE2}
\chi (r) = r + \int\limits_0^a {K(r,t)} \chi (t)dt,
\end{equation}
whose solution enables us to calculate $\varphi_1(r)$.
However, due to the complexity of the kernel $K(r,t)$, a straightforward analytical solution of this equation seems unlikely. On the other hand, by means of the $\psi(r)$ obtained with the second mode, from the third equation of the first mode (see table~{\ref{tab:disc}}) we get
\begin{equation*}
\varphi (r) = S_{1/2, - 1/2} \psi (r) = \frac{{H_0 }}{{\pi ^{3/2} }}\left( {\frac{{2ar}}{{a^2  - r^2 }} - \ln \frac{{a - r}}{{a + r}}} \right) = \frac{{H_0 }}{{\pi ^{3/2} }}\frac{d}{{dr}}\left( {r\ln \frac{{a + r}}{{a - r}}} \right) \quad \quad (r \in I_1)
\end{equation*}
from which we obtain {\it a posteriori} the analytical solution of~(\ref{eq:f2_disc})
\begin{equation}
\chi (r) = \frac{2}{\pi }\sqrt {a^2  - r^2 } \left( {\frac{{ar}}{{a^2  - r^2 }} + \tanh ^{ - 1} r/a} \right).
\end{equation}
Straight numerical tests show that this function verifies equation~(\ref{eq:FredholmIE2}). 
The analytical difficulties resulting from the choice of the first mode strikingly differ from the elementary way of finding the solutions with the other two modes. Therefore the correct choice of the mode is imperative for an easy solution (if any) of the problem.

\section{Superconducting disc in uniform magnetic field}\label{sec:sc_disc_field}
\begin{figure}
\centering
\includegraphics[width=0.45\textwidth] {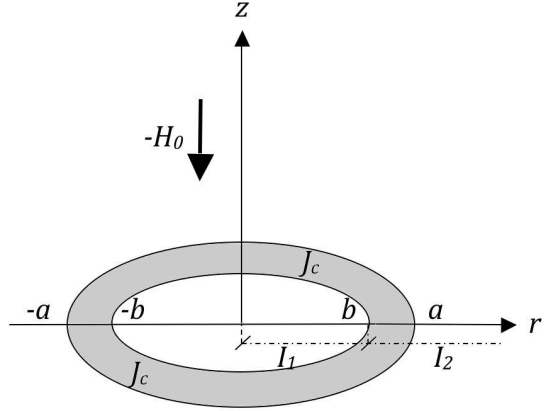}
\caption{Superconducting disc in uniform magnetic field.}
\label{fig:disc_sc}
\end{figure}

We shall now extend the same problem to the case of a superconducting disc in the critical state. According to the critical state hypotheses, the current density can only reach $J_c$ as maximum value and in the regions where this happens the magnetic field penetrates the superconductor. On the contrary, in the regions where the current density is smaller than $J_c$ the magnetic field is zero (null field zone). In the particular case of a disc of radius $a$ subjected to a uniform magnetic field $-H_0$, in the circular region $0<r<b$ the magnetic field  is null and the current density is smaller than $J_c$. In the annular region $b<r<a$ there is a constant current density equal to $J_c$ and the radial magnetic field is $J_c/2$ -- see figure~\ref{fig:disc_sc}. Outside the disc the radial magnetic field component is null by symmetry.

We consider again equations~(\ref{eq:S00psi}) and (\ref{eq:S120psi}) of the previous section and adapt them to the present case.
We divide the $r$-axis in the two intervals $I_1 :(0,b) \cup I_2 :(b,\infty )$ and define $f_1(r)=H_0$ (as before); now $g_2$ is no longer null everywhere, but according to the critical state description it is
\begin{align*}
g_2 (r) = \left\{ {\begin{array}{*{20}c}
   {{\textstyle{1 \over 2}}J_c \quad \quad (b < r < a)}  \\
   {0\quad \quad (a < r < \infty)}  \\
\end{array}} \right.
\end{align*}
in order to account for the partial penetration of the field in the disc.
\subsection{Second mode}
Developing the equations of the second mode (second column in table~\ref{tab:disc}), in the intervals $I_1$ and $I_2$ we obtain the following four equations
\begin{align}
\label{eq:f1_disc_sc}
& H_0 =\left( {\begin{array}{*{20}c}
   r  \\
   0  \\
\end{array}} \right)I_{1/2, - 1/2} \varphi _1 (r)
& \quad \quad (r \in I_1)\\
\label{eq:f2_disc_sc}
& f_2 (r) = \left( {\begin{array}{*{20}c}
   b  \\
   0  \\
\end{array}} \right)I_{1/2, - 1/2} \varphi _1 (r) + \left( {\begin{array}{*{20}c}
   r  \\
   b  \\
\end{array}} \right)I_{1/2, - 1/2} \varphi _2 (r) 
& \quad \quad (r \in I_2) \\
\label{eq:g1_disc_sc}
& g_1 (r) = \left( {\begin{array}{*{20}c}
   b  \\
   r  \\
\end{array}} \right)K_{1/2, - 1/2} \varphi _1 (r) + \left( {\begin{array}{*{20}c}
   \infty   \\
   b  \\
\end{array}} \right)K_{1/2, - 1/2} \varphi _2 (r)
& \quad \quad (r \in I_1)\\
\label{eq:g2_disc_sc}
&
g_2 (r) = \left( {\begin{array}{*{20}c}
   a  \\
   r  \\
\end{array}} \right)K_{1/2, - 1/2} \varphi _2 (r) & (r \in I_2).
\end{align}
Inverting (\ref{eq:f1_disc_sc}) we have
\begin{align*}
\varphi _1 (r) = \left( {\begin{array}{*{20}c}
   r  \\
   0  \\
\end{array}} \right)I_{0,1/2} (H_0 ) = \frac{2}{{\sqrt \pi  }}H_0 
& \quad \quad (r \in I_1),
\end{align*}
and inverting (\ref{eq:g2_disc_sc})
\begin{align*}
\varphi _2 (r) = \left( {\begin{array}{*{20}c}
   a  \\
   r  \\
\end{array}} \right)K_{0,1/2} g_2 (r) = \left\{ {\begin{array}{*{20}c}
   {({\textstyle{1 \over 2}}J_c )\frac{2}{{\sqrt \pi  }}\cosh ^{ - 1} \frac{a}{r}\quad (b < r < a)}  \\
   {0\quad (a < r < \infty )}  \\
\end{array}} \right..
\end{align*}
Differently from the case of the perfect conducting  disc, the $\varphi_2$ part is different from zero in the $I_2$ subinterval $b<r<a$ (annular region). In the interval $I_1$ the sheet current will be, according to~(\ref{eq:g1_disc_sc})
\begin{align}
\nonumber
J_{\phi 1} (r) &= 2g_1 = 2K_{1/2, - 1/2} \varphi  = 2\left( {\begin{array}{*{20}c}
   b  \\
   r  \\
\end{array}} \right)K_{1/2, - 1/2} \varphi _1 (r) + 2\left( {\begin{array}{*{20}c}
   a  \\
   b  \\
\end{array}} \right)K_{1/2, - 1/2} \varphi _2 (r)
\\
\nonumber
&= - \frac{4}{\pi }\frac{d}{{dr}}\left\{ {H_0 \int\limits_r^b {\frac{t}{{\sqrt {t^2  - r^2 } }}dt}  + {\textstyle{1 \over 2}}J_c \int\limits_b^a {\frac{t}{{\sqrt {t^2  - r^2 } }}\cosh \frac{a}{t}dt} } \right\}
\\
\label{eq:Jphi1_discsc}
&= \frac{4}{\pi }\left( {H_0  - {\textstyle{1 \over 2}}J_c \cosh ^{ - 1} \frac{a}{b}} \right)\frac{r}{{\sqrt {b^2  - r^2 } }} + \frac{4}{\pi }\left( {{\textstyle{1 \over 2}}J_c } \right)\tan ^{ - 1} \frac{r}{a}\sqrt {\frac{{a^2  - b^2 }}{{b^2  - r^2 }}} \quad \quad (0<r<b). 
\end{align}
From the first term we see that $J_{\phi 1}$ has an infinite discontinuity in $r=b$, which is physically unsound, since it is an internal point of the disc. In order to remove it, we request that the term inside the bracket is null, i.e.
\begin{equation}\label{eq:b}
b = \frac{a}{{\cosh (H_0 /H_c)}},
\end{equation}
where $H_c={\textstyle{1 \over 2}}J_c $.
We see then that the removal of the infinity fixes the hitherto arbitrary value of $b$, the inner radius of the annular region of constant sheet current. Therefore the sheet current becomes
\begin{equation}\label{eq:Jphir}
J_\phi  (r) = \frac{2}{\pi }J_c \tan ^{ - 1} \frac{r}{a}\sqrt {\frac{{a^2  - b^2 }}{{b^2  - r^2 }}} \quad \quad (0<r<b).
\end{equation}
We have in this way found the result of~\cite{J:1993:Mikheenko93}.
In that work the authors obtained the same result by means of a superposition of the allowed solutions and after having produced a Volterra integral equation solved by Mellin transform.
Unfortunately, no details of the mathematical derivation are provided and only the final results (which coincide with equations (\ref{eq:b}) and (\ref{eq:Jphir})) are given. Consequently a full comparison of the analytical work requested to solve the problem is impossible.
Here, with the powerful formalism of the EK operators, the same results are obtained in an effortless manner, with a little modification of the easy case of the perfectly conducting disc, which only requires the solution of the two elementary integrals present in~(\ref{eq:Jphi1_discsc}).

\begin{figure}
\centering
\includegraphics[width=0.45\textwidth] {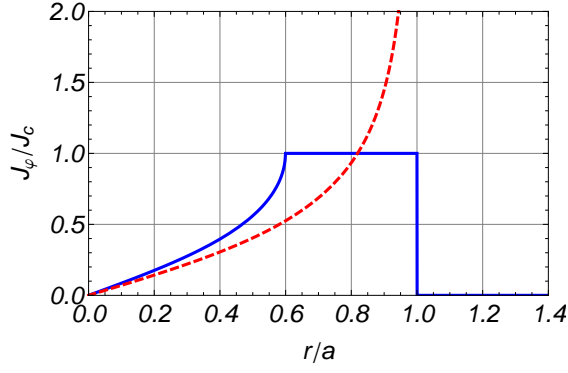}
\caption{Radial sheet current distribution in a disc subjected to an external field: superconductor (continuous line) and perfect conductor (dashed line).}
\label{fig:J_disc}
\end{figure}

\begin{figure}
\centering
\includegraphics[width=0.45\textwidth] {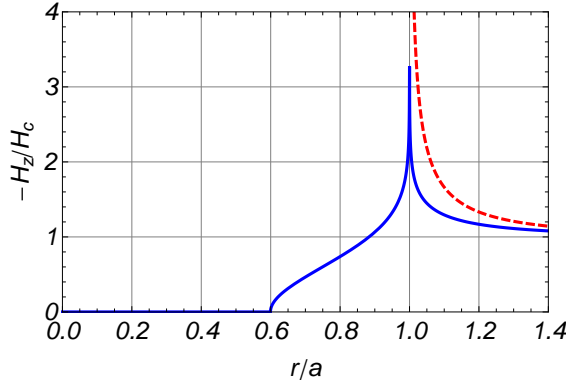}
\caption{Magnetic field distribution in a disc subjected to an external field: superconductor (continuous line) and perfect conductor (dashed line).}
\label{fig:H_disc}
\end{figure}

We can calculate the magnetic field $H_z$ in the annulus $(b<r<a)$. Using~(\ref{eq:g1_disc_sc}) we have
\begin{align}
\nonumber
H_{z} (r) &=  - H_0  + f_2 (r) =  - H_0  + \left( {\begin{array}{*{20}c}
   b  \\
   0  \\
\end{array}} \right)I_{1/2, - 1/2} \varphi _1 (r) + \left( {\begin{array}{*{20}c}
   r  \\
   b  \\
\end{array}} \right)I_{1/2, - 1/2} \varphi _2 (r) \\
\nonumber 
\label{eq:Hzr_discsc1}
&= - H_0  + \frac{1}{{\sqrt \pi  }}\frac{1}{r}\frac{d}{{dr}}\left\{ {\int\limits_0^b {\frac{{t^2 }}{{\sqrt {r^2  - t^2 } }}\varphi _1 (t)dt + \int\limits_b^r {\frac{{t^2 }}{{\sqrt {r^2  - t^2 } }}\varphi _2 (t)dt} } } \right\} \\
&= - H_0  + \frac{2}{\pi }H_0 \left( {\sin ^{ - 1} \frac{b}{r} - \frac{b}{{\sqrt {r^2  - b^2 } }}} \right) + \frac{1}{\pi }J_c \frac{1}{r}\frac{d}{{dr}}\int\limits_b^r {\frac{{t^2 \cosh ^{ - 1} (a/t)}}{{\sqrt {r^2  - t^2 } }}dt}. 
\end{align}
The last integral appears to be analytically unsolvable and we shall be satisfied with a numerical evaluation. Out of the disc, the magnetic field will be
\begin{align}
\nonumber
H_z (r) &=  - H_0  + f_2 (r) =  - H_0  + \left( {\begin{array}{*{20}c}
   b  \\
   0  \\
\end{array}} \right)I_{1/2, - 1/2} \varphi _1 (r) + \left( {\begin{array}{*{20}c}
   a  \\
   b  \\
\end{array}} \right)I_{1/2, - 1/2} (J_c ) \\
\label{eq:Hzr_discsc2}
&=- H_0  + \frac{2}{\pi }H_0 \left( {\sin ^{ - 1} \frac{b}{r} - \frac{b}{{\sqrt {r^2  - b^2 } }}} \right) + \frac{1}{\pi }J_c \int\limits_b^a {\frac{{t^2 \cosh ^{ - 1} (a/t)}}{{(r^2  - t^2 )^{3/2} }}dt}.
\end{align}
From equations (\ref{eq:Hzr_discsc1}) and (\ref{eq:Hzr_discsc2}) we see that the field is the sum of the field of a perfect disc of radius $b$, see equation (4.6), and of the field generated by the constant sheet current $J_c$ in the annulus $b<r<a$.
The difference between the profiles of the sheet current and the magnetic field for the perfect disc and superconducting disc is shown in figures~\ref{fig:J_disc} and \ref{fig:H_disc}, respectively. 


\begin{figure}
\centering
\includegraphics[width=0.45\textwidth] {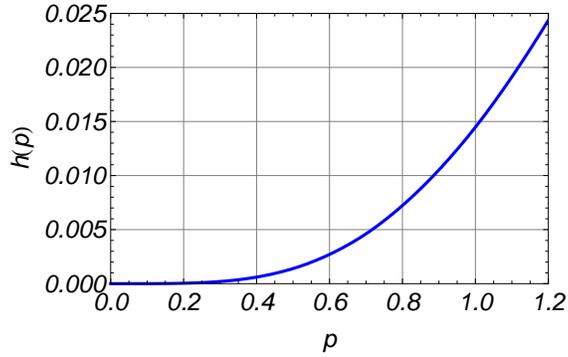}
\caption{Loss function $h(p)$ for a superconducting disc in applied magnetic field.}
\label{fig:hp_disc}
\end{figure}

The ac losses per cycle of period $T$ generated in the disc can be computed with the approach proposed in~\cite{J:1970:Norris70}, which only requires the knowledge of $H_z$ in the annular region at the peak instant
\begin{align*}
L_c  &= \int\limits_0^T {dt} (2\pi )\int\limits_b^a {} JErdr = 2\pi J_c \int\limits_0^T {dt} \int\limits_b^a {E(r)rdr} \\
&= 2\pi \mu _0 J_c \int\limits_0^T {dt} \int\limits_b^a {rdr\left( {\frac{1}{{2\pi r}}\int\limits_b^r {\frac{{dH_z(r',t)}}{{dt}}} (2\pi )r'dr'} \right)} \\
&= 8\pi \mu _0 J_c \int\limits_b^a {(a - r')H_z (r')} r'dr'=L_0h(H_0/H_c),
\end{align*}
where we have defined a reference loss $L_0  = 8\pi \mu _0 J_c^2 a^3$ and $h(p)$ is an adimensional function numerically computed by means of (\ref{eq:Hzr_discsc1}) and shown in figure~\ref{fig:hp_disc}.

\section{Perfectly conducting tape in uniform magnetic field}\label{sec:pc_field}
\begin{figure}
\centering
\includegraphics[width=0.45\textwidth] {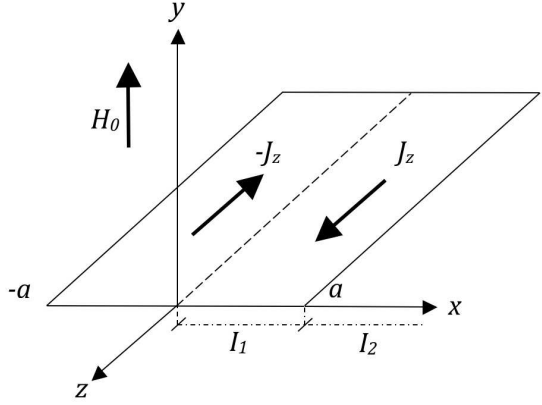}
\caption{Perfectly conducting tape in uniform magnetic field}
\label{fig:tape_pc}
\end{figure}

Our concern here is to derive the dual operator system to determine the current density in the cross-section of a perfectly conducting thin tape of width $2a$ immersed in a uniform perpendicular magnetic field $H_0$. The tape is represented by the segment $(-a,a)$ on the $x$-axis and the field is applied in the $y$-direction, as schematically represented in figure~\ref{fig:tape_pc}. 

If the tape is infinitely long in the $z$ direction and insulated, the external magnetic field induces opposite currents in the two halves $(0,a)$ and $(-a,0)$, which means that the current density distribution is an odd function of $x$, i.e. $J(-x)=-J(x)$, so that it is sufficient to consider only the positive $x$-axis. Given the anti-symmetry of the currents, the magnetic field generated by them can be expressed in the following integral form (see equations~(\ref{eq:Hx_appB2}-\ref{eq:Hy_appB2}) in appendix B)
\begin{align*}
H_x (x) &=  - \,(\pi x/2)^{1/2} \,S_{1/4,0} \psi (x)
\\
H_y (x) &=  - \,(\pi x/2)^{1/2} \,S_{ - 1/4,0} \psi (x).
\end{align*}
By applying a field ramp of amplitude $H_0$ in the $y$-direction, currents are induced in the tape to oppose the inducing field. If the tape is a perfect conductor, those currents will null the external field on the conductor's surface at each instant and, when the external field reaches its final value $H_0$, the induced sheet current will reach a final spatial distribution so that the total magnetic field in the interval $(-a,a)$ is null. Therefore in the plane of the tape ($y=0$) we will have
\begin{align*}
H_y (x) &=  - (\pi x/2)^{1/2} \,S_{ - 1/4,0} \psi (x) =  - H_0 
& (0<x<a)
\\
H_x (x) &=  - (\pi x/2)^{1/2} \,S_{1/4,0} \psi (x) = 0
& (a<x<\infty).
\end{align*}
The second equation tells us that, due to the symmetry, there is no $x$-component of the magnetic field in the plane $y=0$ outside the tape. By dividing the positive $x$-axis in the two complementary intervals $I_1 :(0,a) \cup I_2 :(a,\infty )$, and by defining $f = f_1 \dot  + f_2$ with $f_1 (x) = (\pi x/2)^{ - 1/2} H_0$ and $g = g_1 \dot  + g_2$ with $g_2 (x) = 0$, so that $H_y  =  - (\pi x/2)^{1/2} f$ and $H_x  =  - (\pi x/2)^{1/2} g$, the system becomes
\begin{align}
\tag{i}\label{eq:S-140}
S_{ - 1/4,0} \psi (x) &= f(x) \\
\tag{ii}\label{eq:S140}
S_{1/4,0} \psi (x) &= g(x).
\end{align}
In order to transform these two equations into a couple of equations with the same first member, we can choose between the three modes listed in table~\ref{tab:tape_field}.

\subsection{Second mode}
Since we know the couple $(f_1,g_2)$, our favorite mode will be the second one. The equations of the second column expanded on the two intervals hence will be
\begin{align}
\label{eq:f1_tape_pc}
& f_1 (x) = \left( {\begin{array}{*{20}c}
   x  \\
   0  \\
\end{array}} \right)I_{1/4, - 1/2} \varphi _1 (x)
& \quad \quad (x \in I_1 ) \\
\label{eq:f2_tape_pc}
& f_2 (x) = \left( {\begin{array}{*{20}c}
   a  \\
   0  \\
\end{array}} \right)I_{1/4, - 1/2} \varphi _1 (x) + \left( {\begin{array}{*{20}c}
   x  \\
   a  \\
\end{array}} \right)I_{1/4, - 1/2} \varphi _2 (x)
& \quad \quad (x \in I_2 ) \\
\label{eq:g1_tape_pc}
& g_1(x)  = \left( {\begin{array}{*{20}c}
   a  \\
   x  \\
\end{array}} \right)K_{1/4, - 1/2} \varphi _1 (x) + \left( {\begin{array}{*{20}c}
   \infty   \\
   x  \\
\end{array}} \right)K_{1/4, - 1/2} \varphi _2 (x)
& \quad \quad (x \in I_1 ) \\
\label{eq:g2_tape_pc}
& 0 = \left( {\begin{array}{*{20}c}
   \infty   \\
   x  \\
\end{array}} \right)K_{1/4, - 1/2} \varphi _2 (x)
& (x \in I_2 ).
\end{align}
From (\ref{eq:g2_tape_pc}) we obviously derive that $\varphi _2  = 0$, whereas from (\ref{eq:f1_tape_pc}), by inversion
\begin{align}\label{eq:phi1tapep}
\nonumber
\varphi _1 (x) &= \left( {\begin{array}{*{20}c}
   x  \\
   0  \\
\end{array}} \right)I_{ - 1/4,1/2} f_1 (x) = \frac{2}{{\sqrt \pi  }}x^{ - 1/2} \int\limits_0^x {\frac{{t^{1/2} }}{{\sqrt {x^2  - t^2 } }}f_1 (t)dt} 
\\
\nonumber
&= \sqrt 2 H_0 x^{ - 1/2} \frac{2}{\pi }\int\limits_0^x {\frac{{dt}}{{\sqrt {x^2  - t^2 } }} = } \sqrt 2 H_0 x^{ - 1/2} \quad \quad (x \in I_2). 
\end{align}
By means of equation (\ref{eq:Jz_appB2}) of appendix B, the current density is $J_z (x) =  - 2H_x (x) = 2(\pi x/2)^{1/2} g_1 (x)$, so that using~(\ref{eq:g1_tape_pc}) we can write
\begin{eqnarray}
J_z (x) &=& 2(\pi x/2)^{1/2} \left( {\begin{array}{*{20}c}
   a  \\
   x  \\
\end{array}} \right)K_{1/4, - 1/2} \varphi _1 (x)
\nonumber \\
&=& 2(\pi x/2)^{1/2} \left( { - \frac{{x^{ - 1/2} }}{{\sqrt \pi  }}\frac{d}{{dx}}\int\limits_x^a {\frac{{t^{3/2} }}{{\sqrt {t^2  - x^2 } }}} \varphi _1 (t)dt} \right)
\nonumber \\
&=& - 2H_0 \frac{d}{{dx}}\int\limits_x^a {\frac{{t^{} }}{{\sqrt {t^2  - x^2 } }}} dt = 2H_0 \frac{x}{{\sqrt {a^2  - x^2 } }}.
\label{eq:Jzx_tapepc_II}
\end{eqnarray}
It can be noted that the solution is obtained by means of two trivial integrals and does not need the explicit computation of the Henkel transforms. Apart from a numerical coefficient, this current profile is identical to that found for the perfectly conducting disc -- see equation~(\ref{eq:disc_Jz}).

In the plane of the tape ($y=0$), the total magnetic field has only the transversal component $H_y$, which is the sum of the applied and the induced field
\begin{equation*}
H_y (x) = H_0  - (\pi x/2)^{1/2} f_2 (x).
\end{equation*}
By using (\ref{eq:f2_tape_pc}), the magnetic field becomes
\begin{equation*}
\label{eq:Hy_tape_II}
H_y (x) =  H_0  - H_0 \frac{d}{{dx}}\int\limits_0^a {\frac{t}{{\sqrt {x^2  - t^2 } }}dt = } H_0 \frac{x}{{\sqrt {x^2  - a^2 } }} \quad \quad (x \in I_2).
\end{equation*}
Since we know $\varphi=\varphi_1$  on all real positive axis, we can also obtain the $\psi$  function from the third equation of the second mode
\begin{align}
\label{eq:psi_discp}
\psi (x) = S_{1/4, - 1/2} \varphi _1  = \sqrt {\frac{x}{2}} \int\limits_0^a {t^{3/2} \varphi _1 (t)J_0 (x,t)dt = \frac{{aH_0 }}{{\sqrt {2x} }}J_1 (ax)} 
\quad \quad(0 < x < \infty ).
\end{align}
\subsection{Third mode}
We now try to use the third mode. Making the second equation explicit in $I_1$ and using the fact that $g_2=0$, we shall have
\begin{equation}
\nonumber
\left( {\begin{array}{*{20}c}
   x  \\
   0  \\
\end{array}} \right)I_{ - 1/4,1/2} f_1 (x) = \left( {\begin{array}{*{20}c}
   a  \\
   x  \\
\end{array}} \right)K_{ - 1/4,1/2} g_1 (x)\quad \quad (x \in I_1 ),
\end{equation}
from which we can immediately derive $g_1$ by inversion
\begin{align}
\nonumber
g_1 (x) &= \left( {\begin{array}{*{20}c}
   a  \\
   x  \\
\end{array}} \right)K_{1/4, - 1/2} \left( {\begin{array}{*{20}c}
   x  \\
   0  \\
\end{array}} \right)I_{ - 1/4,1/2} f_1 (x) \\
\nonumber
&= \sqrt 2 H_0 \left( {\begin{array}{*{20}c}
   a  \\
   x  \\
\end{array}} \right)K_{1/4, - 1/2} (x^{ - 1/2} ) =
 \frac{H_0{(2 x/\pi )}^{1/2} }{{\sqrt {a^2  - x^2 } }}.
\end{align}
The current distribution is immediately obtained as
\begin{equation*}
J_z (x) =  - 2H_x (x) = 2(\pi x/2)^{1/2} g_1 (x) = 2H_0 \frac{x}{{\sqrt {a^2  - x^2 } }},
\end{equation*}
which coincides with the result (\ref{eq:Jzx_tapepc_II}) previously found with the second mode. Developing the same equation on $I_2$  we obtain
\begin{equation*}
\left( {\begin{array}{*{20}c}
   a  \\
   0  \\
\end{array}} \right)I_{ - 1/4,1/2} f_1  + \left( {\begin{array}{*{20}c}
   a  \\
   x  \\
\end{array}} \right)I_{ - 1/4,1/2} f_2^{}  = 0 \quad \quad (x \in I_2)
\end{equation*}
from which we directly derive, using the double operator $L_{\eta ,\alpha }$ (see equation~(\ref{eq:double_Ina}) in appendix A)
\begin{align*}
f_2  &=  - \left( {\begin{array}{*{20}c}
   x  \\
   a  \\
\end{array}} \right)I_{ - 1/4,1/2}^{ - 1} \left( {\begin{array}{*{20}c}
   a  \\
   0  \\
\end{array}} \right)I_{ - 1/4,1/2} f_1^{}  =  - \left( {\begin{array}{*{20}c}
   x & a  \\
   a & 0  \\
\end{array}} \right)L_{ - 1/4,1/2} f_1 \\
& =
 - \frac{2}{\pi }\sqrt {\frac{x}{{x^2  - a^2 }}} \int\limits_0^a {\frac{{\sqrt {a^2  - t^2 } }}{{x^2  - t^2 }}t^{1/2} (\pi t/2)^{ - 1/2} H_0 dt}  \\
& =
(\pi /2)^{ - 1/2} H_0 \sqrt {\frac{x}{{x^2  - a^2 }}} \left( {\frac{{\sqrt {x^2  - a^2 } }}{x} - 1} \right).
\end{align*}
From the expression for $f_2$ we can derive the total magnetic field in $I_2$ as
\begin{equation*}
H_y (x) = H_0  - (\pi x/2)^{1/2} f_2 (x)=H_0 \frac{x}{{\sqrt {x^2  - a^2 } }},
\end{equation*}
which coincides with the expression~(\ref{eq:Hy_tape_II}) found with the second mode.
It has to be noted that, using the third mode, we can determine the physical quantities of interest just by means of one integral, without having to compute the intermediate function $\varphi$, as in the case of the second mode.

\subsection{First mode}
For completeness we try also the first mode. Making the two equations explicit in the two intervals, we shall have the following system
\begin{align}
\label{eq:f1_tape_pc1}
& f_1  = \left( {\begin{array}{*{20}c}
   a  \\
   x  \\
\end{array}} \right)K_{ - 1/4,1/2} \varphi _1  + \left( {\begin{array}{*{20}c}
   \infty   \\
   x  \\
\end{array}} \right)K_{ - 1/4,1/2} \varphi _2 
& (x \in I_1 )
\\
\label{eq:f2_tape_pc1}
& f_2  = \left( {\begin{array}{*{20}c}
   \infty   \\
   x  \\
\end{array}} \right)K_{ - 1/4,1/2} \varphi _2 
& (x \in I_2 )
\\
\label{eq:g1_tape_pc1}
& g_1  = \left( {\begin{array}{*{20}c}
   x  \\
   0  \\
\end{array}} \right)I_{ - 1/4,1/2} \varphi _1 
& (x \in I_1 )
\\
\label{eq:g2_tape_pc1}
& 0 = \left( {\begin{array}{*{20}c}
   a  \\
   0  \\
\end{array}} \right)I_{ - 1/4,1/2} \varphi _1  + \left( {\begin{array}{*{20}c}
   x  \\
   a  \\
\end{array}} \right)I_{ - 1/4,1/2} \varphi _2 
& (x \in I_2 ).
\end{align}
From (\ref{eq:f1_tape_pc1}) we derive 
\begin{align*}
\varphi _1 (x) &=
 - \left( {\begin{array}{*{20}c}
   a & \infty   \\
   x & a  \\
\end{array}} \right)M_{ - 1/4,1/2} \varphi _2 (x) + \left( {\begin{array}{*{20}c}
   a  \\
   x  \\
\end{array}} \right)K_{1/4, - 1/2}^{} f_1 (x),
\end{align*}
whereas from (\ref{eq:g2_tape_pc1})
\begin{equation*}
\varphi _2 (x) =  - \left( {\begin{array}{*{20}c}
   x & a  \\
   a & 0  \\
\end{array}} \right)L_{ - 1/4,1/2} \varphi _1 (x).
\end{equation*}
Eliminating $\varphi_2$, we obtain the Fredholm integral equation of the second kind for $\varphi_1$
\begin{equation*}
\varphi _1  = \left( {\begin{array}{*{20}c}
   a & \infty   \\
   x & a  \\
\end{array}} \right)M_{ - 1/4,1/2} \left( {\begin{array}{*{20}c}
   x & a  \\
   a & 0  \\
\end{array}} \right)L_{ - 1/4,1/2} \varphi _1  + \left( {\begin{array}{*{20}c}
   a  \\
   x  \\
\end{array}} \right)K_{1/4, - 1/2} f_1 
\end{equation*}
or, explicitly, by means of equations (\ref{eq:double_Ina}-\ref{eq:double_Kna}) of appendix A
\begin{equation*}
\varphi _1 (x) = \frac{{\sqrt x }}{{\sqrt {a^2  - x^2 } }}\int\limits_0^a {K(x,t)} \sqrt {a^2  - t^2 } \sqrt t \varphi _1 (t)dt + H_0 \frac{{\sqrt 2 }}{\pi }\frac{{\sqrt x }}{{\sqrt {a^2  - x^2 } }},
\end{equation*}
where the kernel is
\begin{equation*}
K(x,t) = \frac{2}{{\pi ^2 }}\frac{{\ln (a^2  - x^2 ) - \ln (a^2  - t^2 )}}{{t^2  - x^2 }}.
\end{equation*}
Multiplying both sides by $\pi \sqrt {a^2  - x^2 } /(H_0 \sqrt {2x} )$
and posing $\omega (x) = \varphi _1 (x)\pi \sqrt {a^2  - x^2 } /(H_0 \sqrt {2x} )$, we obtain the simpler integral equation
\begin{equation}\label{eq:omega}
\omega (x) = \int\limits_0^a {t\,K(x,t)} \omega (t)dt + 1.
\end{equation}
The direct solution of this integral equation appears to be very hard; 
nevertheless, in order to test the correctness of the first mode, we deduced $\varphi$ by inverting the third equation, i.e. $\varphi  = S_{1/4, - 1/2} \psi$, and, using~(\ref{eq:psi_discp}) obtained from the second mode
\begin{equation*}
\varphi _1 (x) = aH_0 \sqrt {\frac{2}{x}} \int\limits_0^\infty  t J_1 (at)J_0 (x,t)dt = \frac{{aH_0 }}{\pi }\sqrt {2x} \frac{{{\rm{E}}(x/a)}}{{a^2  - x^2 }} \quad \quad (x<a)
\end{equation*}
where ${\rm E}(k)$ is the complete elliptic integral of second type. From this we have
\begin{equation*}
\omega (x) = \frac{{a^2 }}{{\sqrt {a^2  - x^2 } }}{\rm{E}}(x/a).
\end{equation*}
Numerical tests have confirmed that $\omega(x)$ is the solution of~(\ref{eq:omega}). The current density will be $J_z (x) =  - 2H_r (x) = 2(\pi x/2)^{1/2} g_1 (x)$ and, by~(\ref{eq:g1_tape_pc1}),
\begin{align*}
J_z (x) = 2(\pi x/2)^{1/2} \left( {\begin{array}{*{20}c}
   x  \\
   0  \\
\end{array}} \right)I_{ - 1/4,1/2} \varphi _1 (x) = 2H_0 \frac{2}{\pi }a\int\limits_0^x {\frac{{t\,{\rm{E}}(t/a)}}{{(a^2  - t^2 )\sqrt {x^2  - t^2 } }}dt}  = 2H_0 \frac{x}{{\sqrt {a^2  - x^2 } }}.
\end{align*}
We have obtained again the previous result (\ref{eq:Jzx_tapepc_II}), but with much more analytical difficulties. The `wrong' mode still gives the correct results but  at the price of renouncing any easy operator inversion. Actually, in cases like this we cannot anymore rely on the disentangling properties of the EK operators, but we have to solve Fredholm integral equations with very complex singular kernels, often a difficult task that can be managed mostly only by means of numerical methods.

\section{Perfectly conducting tape with transport current}
\label{sec:tape_pc_curr}
We now consider the case of a perfectly conducting tape with transport current (self-field). By symmetry the current density is an even function of $x$, i.e.  $J(-x)=J(x)$, which can null the transverse magnetic field $H_y$ inside the whole tape at each time instant. By symmetry also the magnetic field component $H_x$ is null in the whole plane $y=0$ out of the tape. Therefore, having divided the positive real axis as in the preceding section, we can state the following mixed conditions
\begin{align*}
H_y (x) &= 0\quad \quad (x \in I_1 )
\\
H_x (x) &= 0\quad \quad (x \in I_2 ).
\end{align*}
Utilizing (\ref{eq:Hx_appB}) and (\ref{eq:Hy_appB}) of appendix B (currents in the same direction) we can rewrite them as the following dual operator system
\begin{align}
\tag{i}
\label{eq:S-140_tape_pc_curr}
& S_{ - 1/4,0} \psi (x) =  - (\pi x/2)^{ - 1/2} H_x (x) = g(x)\dot  + g_2 (x)
\\
\tag{ii}
\label{eq:S140_tape_pc_curr}
& S_{1/4,0} \psi (x) = (\pi x/2)^{ - 1/2} H_y (x) = f_1 (x)\dot  + f_2 (x),
\end{align} 
where we have $f_1=0$ and $g_2=0$, and the unknown functions $f_2$ and $g_1$ are to be found by using one of the three modes listed in table~\ref{tab:tape_curr}.

The application of the second mode is useless, because both known part functions are null and, as a consequence, we obtain the trivial solution $J=0$. The first mode leads to a Fredholm integral equation of the second type of difficult solution. The third mode results to be the most viable option. 

\subsection{Third mode}
Making the second equation for the third mode (see table~\ref{tab:tape_curr}) explicit in $I_1$, we obtain the following integral equation
\begin{align*}
\left( {\begin{array}{*{20}c}
   x  \\
   0  \\
\end{array}} \right)I_{1/4, - 1/2} f_1  = \left( {\begin{array}{*{20}c}
   a  \\
   x  \\
\end{array}} \right)K_{1/4, - 1/2} g_1  + \left( {\begin{array}{*{20}c}
   \infty   \\
   a  \\
\end{array}} \right)K_{1/4, - 1/2} g_2 
\end{align*}
and, since $f_1$ and $g_2$ are null,
\begin{align*}
0 = \left( {\begin{array}{*{20}c}
   a  \\
   x  \\
\end{array}} \right)K_{1/4, - 1/2} g_1 (x) =  - \frac{1}{{\sqrt \pi  }}\frac{1}{{\sqrt x }}\frac{d}{{dx}}\int\limits_x^a {\frac{{t^{3/2} g_1 (t)}}{{\sqrt {t^2  - x^2 } }}dt}. 
\end{align*}
This equation entails
\begin{align*}
\int\limits_x^a {\frac{{t^{3/2} g_1 (t)}}{{\sqrt {t^2  - x^2 } }}dt}  = C.
\end{align*}
By defining $\gamma (t) = t^{3/2} g_1 (x)$, the equation can be expressed (see table~\ref{tab:A1} in appendix A) in the operator form as follows
\begin{align*}
\int\limits_x^a {\frac{{\gamma (t)}}{{\sqrt {t^2  - x^2 } }}dt}  = K_{0,1/2} \gamma (x) = C,
\end{align*}
so that it can be immediately solved by inversion
\begin{align*}
\gamma (x) = \left( {\begin{array}{*{20}c}
   a  \\
   x  \\
\end{array}} \right)K_{1/2, - 1/2} C = C\frac{x}{{\sqrt {a^2  - x^2 } }}
\end{align*}
and hence $g_1 (x) = \frac{C}{{\sqrt x \sqrt {a^2  - x^2 } }}$.
According to~(\ref{eq:Jz_appB}) of appendix B, the current density is
\begin{align*}
J_z (x) = 2(\pi x/2)^{1/2} g_1  = (2\pi )^{1/2} \frac{C}{{\sqrt {a^2  - x^2 } }}.
\end{align*}
The constant $C$ is determined by the current transported by the tape $I_0  = \int\limits_{ - a}^a {J_z (x)dx}$, from which $C = 2^{ - 1/2} \pi ^{ - 3/2} I_0$.
Then we finally obtain the well known result
\begin{equation}
J_z (x) = \frac{{I_0 }}{{\pi \sqrt {a^2  - x^2 } }}.
\end{equation}
From equation (\ref{eq:S140_tape_pc_curr}) we can immediately obtain the transverse magnetic field on the tape plane $H_y (x) = (\pi x/2)^{1/2} S_{1/4,0} \psi (x)$. Since we know $g(x)$ from (\ref{eq:S-140_tape_pc_curr}) we have
\begin{equation}
\psi (x) = \left( {\begin{array}{*{20}c}
   a  \\
   0  \\
\end{array}} \right)S_{ - 1/4,0} g_1 (x) = (2\pi )^{ - 1} x^{ - 1/2} J_0 (ax)
\end{equation}
so that, knowing $\psi$, from~(\ref{eq:S140_tape_pc_curr}) we can obtain (by the Weber-Shafheitling discontinuous integral)
\begin{align}
\label{eq:Hyx}
H_y (x) = (\pi x/2)^{1/2} (2\pi )^{ - 1} \int\limits_0^\infty  {t^{1/2} J_{1/2} (xt)J_0 (at)dt} 
&= \left\{ {\begin{array}{*{20}c}
   {0\quad \quad (0 < x < a)}  \\
   {\frac{{I_0 }}{{2\pi \sqrt {x^2  - a^2 } }}\quad (a < x < \infty)}  \\
\end{array}} \right..
\end{align}
Note that we can also derive the magnetic field avoiding the intermediate function $\psi(x)$.
From (\ref{eq:S140_tape_pc_curr}) we have that in the interval $I_2$ $H_y (x) = (\pi x/2)^{1/2} f_2 (x)$ . We can derive $f_2$  from the first equation of the third mode (see table~\ref{tab:tape_curr})
\begin{equation*}
f_2 (x) = \left( {\begin{array}{*{20}c}
   \infty   \\
   x  \\
\end{array}} \right)K_{1/4, - 1/2} \left( {\begin{array}{*{20}c}
   a  \\
   0  \\
\end{array}} \right)I_{ - 1/4,1/2} g_1 (x)
=
- C\frac{2}{\pi }\frac{1}{{\sqrt x }}\frac{d}{{dx}}\int\limits_x^\infty  {\frac{{K(a/t)}}{{\sqrt {x^2  - t^2 } }}} \,dt.
\end{equation*}
Applying the derivative to the by part integration we obtain
\begin{equation*}
f_2 (x) = C\frac{2}{\pi }\sqrt x \int\limits_x^\infty  {\frac{{E(a/t)}}{{\left( {t^2  - a^2 } \right)\sqrt {x^2  - t^2 } }}} \,dt
=
C\frac{1}{{\sqrt x \sqrt {x^2  - a^2 } }}
\end{equation*}
and the expression for $H_y(x)$ coincides with (\ref{eq:Hyx}).
%
\section{Superconducting tape in the critical state with transport current}
\label{sec:tape_sc_curr}
We consider an infinitely long superconducting tape of width $2a$ with transport current $I_t$ flowing in the plane $y=0$ and with cross-section represented by the segment $(-a,a)$ on the $x$-axis.
According to the assumptions of the critical state, the sheet current is constant and equal to $J_c$ in two lateral bands $(-a,-b)$ and $(b,a)$, where there is a tangential magnetic field $H_x=-J_c/2$. In the central region $(-b,b)$ the field $H_y$ (perpendicular to the tape's plane) is null and so is $H_x$ outside the tape (by symmetry). Given the symmetry of the problem, $J(-x)=J(x)$, we can consider only the $x>0$ region. 
Since $b$ is the critical point, the definition of the two intervals $I_1 :(0,b) \cup I_2 :(b,\infty )$ is quite natural. The mixed boundary conditions are the same as in the case of the perfectly conducting tape, see section~\ref{sec:tape_pc_curr}
\begin{align*}
& S_{ - 1/4,0} \psi (x) =  - (\pi x/2)^{ - 1/2} H_x (x) = g_1(x)\dot  + g_2 (x)
\\
& S_{1/4,0} \psi (x) = (\pi x/2)^{ - 1/2} H_y (x) = f_1 (x)\dot  + f_2 (x),
\end{align*}
where again $f_1(x)=0$, but this time $g_2(x)$ is no longer null, and different from zero in the band $(b<x<a)$, i.e.
\begin{align*}
g_2 (x) = (\pi x/2)^{ - 1/2} \left\{ {\begin{array}{*{20}c}
   {{\textstyle{1 \over 2}}J_c \quad \quad (b < x < a)}  \\
   {0\quad \quad (a < x < \infty )}.  \\
\end{array}} \right.
\end{align*}
The three methods of inversion are the same as those listed in table~\ref{tab:tape_curr}.

\subsection{Second mode}
Since we know the couple $(f_1,g_2)$, we apply the second mode, so that the operator system becomes
\begin{align}\tag{i}\label{eq:fx_norris}
f(x) &= I_{ - \frac{1}{4},\frac{1}{2}} \varphi (x) \\
\tag{ii}\label{gx_norris}
g(x) &= K_{ - \frac{1}{4},\frac{1}{2}} \varphi (x).
\end{align}
Developing them on the two intervals we obtain the system
\begin{align}
\label{eq:f1_tape_sct}
f_1 (x) &= \left( {\begin{array}{*{20}c}
   x  \\
   0  \\
\end{array}} \right)I_{ - \frac{1}{4},\frac{1}{2}} \varphi _1 (x)
& (x \in I_1 )
\\
\label{eq:f2_tape_sct}
f_2 (x) &= \left( {\begin{array}{*{20}c}
   b  \\
   0  \\
\end{array}} \right)I_{ - 1/4,1/2} \varphi _1 (x) + \left( {\begin{array}{*{20}c}
   x  \\
   b  \\
\end{array}} \right)I_{ - 1/4,1/2} \varphi _2 (x) & (x \in I_2)
\\
\label{eq:g1_tape_sct}
g_1 (x) &= \left( {\begin{array}{*{20}c}
   b  \\
   x  \\
\end{array}} \right)K_{ - 1/4,1/2} \varphi _1 (x) + \left( {\begin{array}{*{20}c}
   \infty   \\
   b  \\
\end{array}} \right)K_{ - 1/4,1/2} \varphi _2 (x) & (x \in I_1)
\\
\label{eq:g2_tape_sct}
g_2 (x) &= \left( {\begin{array}{*{20}c}
   \infty   \\
   x  \\
\end{array}} \right)K_{ - \frac{1}{4},\frac{1}{2}} \varphi _2 (x)
& (x \in I_2).
\end{align}
Since $f_1 (x) = 0$, then from (\ref{eq:f1_tape_sct}) we obviously derive $\varphi _1  = 0$ and from~(\ref{eq:g2_tape_sct}) we obtain by inversion
\begin{align*}
\varphi _2 (x) &= \left( {\begin{array}{*{20}c}
   \infty   \\
   x  \\
\end{array}} \right)K_{1/4, - 1/2} g_2 (x) = {\textstyle{1 \over 2}}J_c \left( {\begin{array}{*{20}c}
   a  \\
   x  \\
\end{array}} \right)K_{1/4, - 1/2} (\pi x/2)^{ - 1/2} 
\\
&= \left\{ {\begin{array}{*{20}c}
   {{\textstyle{1 \over 2}}J_c \frac{{\sqrt {2x} }}{{\pi \sqrt {a^2  - x^2 } }}\quad \quad (b < x < a)}  \\
   {0\quad \quad (a < x < \infty ).}  \\
\end{array}} \right .
\end{align*}
Therefore, according to~(\ref{eq:Jz_appB}) of appendix B and to~(\ref{eq:g1_tape_sct}), the sheet current in the interval $I_1$ is
\begin{align}\label{eq:Jz_tape_sc_curr}
\nonumber
J_z (x) &= 2(\pi x/2)^{1/2} g_1 (x) = 2(\pi x/2)^{1/2} \left( {\begin{array}{*{20}c}
   a  \\
   b  \\
\end{array}} \right)K_{ - 1/4,1/2} \phi _2 (x) \\
&= \frac{2}{\pi }J_c \int\limits_a^b {\frac{u}{{\sqrt {u^2  - x^2 } \sqrt {a^2  - u^2 } }}du = } \frac{2}{\pi }J_c \tan ^{ - 1} \sqrt {\frac{{a^2  - b^2 }}{{b^2  - x^2 }}}. 
\end{align}
The value of $b$ is computed by imposing the total transport current
\begin{align*}
I_t  &= \int\limits_{ - a}^a {J_z (x)dx = 2(a - b)J_c  + \frac{2}{\pi }J_c \int\limits_{ - b}^b {\tan ^{ - 1} \sqrt {\frac{{a^2  - b^2 }}{{b^2  - x^2 }}} } } dx \\
&= 2(a - b)J_c  + 2(b - a + \sqrt {a^2  - b^2 } )J_c, 
\end{align*}
from which
\begin{equation}\label{eq:b_tape_sc_curr}
b = a\sqrt {1 - (I_t /2aJ_c )^2 }. 
\end{equation}
In order to compute the ac losses, one needs to know the magnetic field $H_y$ in the interval $(b,a)$. In the sub-interval $(b,a)$ of $I_2$ we have from (\ref{eq:f2_tape_sct}), remembering that $\varphi_1=0$,
\begin{align*}
f_2 (x) &= \left( {\begin{array}{*{20}c}
   x  \\
   b  \\
\end{array}} \right)I_{ - 1/4,1/2} \varphi _2 (x) \\
&= \frac{{J_c }}{\pi }(\pi x/2)^{ - 1/2} \int\limits_b^x {\frac{u}{{\sqrt {x^2  - u^2 } \sqrt {a^2  - u^2 } }}} du
= \frac{{J_c }}{\pi }(\pi x/2)^{ - 1/2} \tanh ^{ - 1} \frac{{\sqrt {x^2  - b^2 } }}{{\sqrt {a^2  - b^2 } }},
\end{align*}
from which

\begin{align}\label{eq:Hy_tape_sc_curr}
H_y (x) = (\pi x/2)^{1/2} \,f_2 (x) = \frac{{J_c }}{\pi }\tanh ^{ - 1} \frac{{\sqrt {x^2  - b^2 } }}{{\sqrt {a^2  - b^2 } }}.
\end{align}
Results (\ref{eq:Jz_tape_sc_curr}), (\ref{eq:b_tape_sc_curr}) and (\ref{eq:Hy_tape_sc_curr}) are the same as those obtained by Norris by means of an {\it ad hoc} conformal transform and image methods; here, on the contrary, we obtained them by means of a general operator scheme applied to a simple two part mixed boundary problem.

By means of the Norris method, the ac losses per cycle can be obtained by integrating the magnetic flux in the two side bands
\begin{equation*}
L_c  = 8\mu _0 J_c \int\limits_b^a {(x - b)H_y (x)dx}. 
\end{equation*}
Using (\ref{eq:b_tape_sc_curr}) and (\ref{eq:Hy_tape_sc_curr}), we can solve this integral analytically, obtaining the well-known formula
\begin{equation}
L_c  = \frac{{I_c^2 \mu _0 }}{\pi }\left[ {(1 - p)\ln (1 - p) + (1 + p)\ln (1 + p) - p^2 } \right],
\end{equation}
where $p=I_t/I_c$.
\subsection{Third mode}
The same solution can be easily obtained also with the third mode by using the second equation 
\begin{equation*}
I_{1/4, - 1/2} f = K_{1/4, - 1/2} g
\end{equation*}
which, once made explicit in the two intervals and since $g_2(x)=0$ for $a<x<\infty$, becomes
\begin{align}\label{eq:x0tapes}
&
\left( {\begin{array}{*{20}c}
   x  \\
   0  \\
\end{array}} \right)I_{1/4, - 1/2} f_1  = \left( {\begin{array}{*{20}c}
   b  \\
   x  \\
\end{array}} \right)K_{1/4, - 1/2} g_1  + \left( {\begin{array}{*{20}c}
   \infty   \\
   b  \\
\end{array}} \right)K_{1/4, - 1/2} g_2
& (x \in I_1) 
\\
\label{eq:b0tapes}
&
\left( {\begin{array}{*{20}c}
   b  \\
   0  \\
\end{array}} \right)I_{1/4, - 1/2} f_1  + \left( {\begin{array}{*{20}c}
   x  \\
   b  \\
\end{array}} \right)I_{1/4, - 1/2} f_2  = \left( {\begin{array}{*{20}c}
   a   \\
   x  \\
\end{array}} \right)K_{1/4, - 1/2} g_2 
& (x \in I_2)
\end{align}
and, since $f_1=0$, from~(\ref{eq:x0tapes}) we obtain in $I_1$, by means of the double operator $M_{\eta,\alpha}$ (see equation~(\ref{eq:double_Kna}) in appendix A)
\begin{align*}
g_1  &=  - \left( {\begin{array}{*{20}c}
   b  \\
   x  \\
\end{array}} \right)K_{1/4,1/2}^{ - 1} \left( {\begin{array}{*{20}c}
   a   \\
   b  \\
\end{array}} \right)K_{1/4, - 1/2} g_2  =  - \left( {\begin{array}{*{20}c}
   b & a   \\
   x & b  \\
\end{array}} \right)M_{1/4, - 1/2} g_2  \\
\nonumber
&= \frac{1}{\pi }J_c (\pi x/2)^{ - 1/2} \sqrt {b^2  - x^2 } \int\limits_b^a {\frac{t}{{\sqrt {t^2  - b^2 } (t^2  - x^2 )}}} dt
 \\
&= J_c \frac{1}{\pi }(\pi x/2)^{ - 1/2} \tan ^{ - 1} \sqrt {\frac{{a^2  - b^2 }}{{b^2  - x^2 }}}  \quad \quad (0<x<b).
\end{align*}
The current density in $I_1$ will be by~(\ref{eq:Jz_appB})
\begin{equation*}
J_z (x) = 2(\pi x/2)^{1/2} g_1 (x) = J_c \frac{2}{\pi }\tan ^{ - 1} \sqrt {\frac{{a^2  - b^2 }}{{b^2  - x^2 }}}  \quad \quad (0<x<b),
\end{equation*}
identical to~(\ref{eq:Jz_tape_sc_curr}). The magnetic field $H_y$ in $I_2$ is given by $H_y (x) = (\pi x/2)^{1/2} \,f_2 (x)$ and $f_2$ can be obtained from~(\ref{eq:b0tapes}), i.e. in the interval $(b<x<a)$
\begin{equation*}
f_2 (x) = \left( {\begin{array}{*{20}c}
   x  \\
   b  \\
\end{array}} \right)I_{ - 1/4,1/2} \left( {\begin{array}{*{20}c}
   a  \\
   x  \\
\end{array}} \right)K_{1/4, - 1/2} g_2  = \frac{1}{\pi }J_c (x\pi /2)^{ - 1/2} \tanh ^{ - 1} \sqrt {\frac{{x^2  - b^2 }}{{a^2  - b^2 }}} , 
\end{equation*}
a result identical to~(\ref{eq:Jz_tape_sc_curr}). We note that with the third mode we avoided using the auxiliary functions $\psi$ and $\varphi$.
%
%
%
%
\section{Superconducting tape in the critical state in uniform magnetic field}
\begin{figure}
\centering
\includegraphics[width=0.45\textwidth] {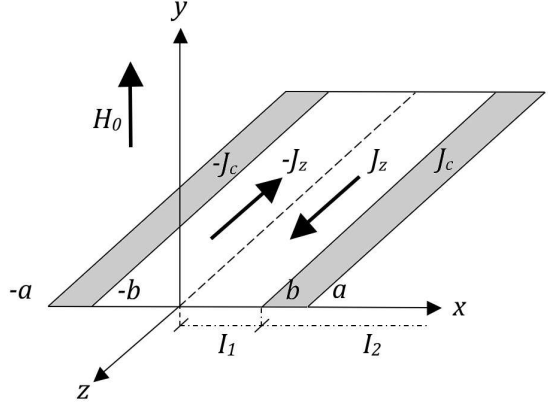}
\caption{Superconducting tape in uniform field}
\label{fig:tape_sc}
\end{figure}

Let us consider again the superconducting tape of the previous section, this time subjected to a uniform magnetic field $H_0$ and with no net transport current. The induced sheet current cannot exceed the critical value $J_c$ and creates a region $(-b,b)$ in the center of the tape with transverse field $H_y=-H_0$, so that the resulting total magnetic field is zero (null field zone), see figure~\ref{fig:tape_sc}. In the lateral bands $(-a,-b)$ and $(b,a)$ the current density assumes the constant critical value $\pm J_c$ and consequently there the tangential magnetic field is $H_x=\mp J_c/2$. Since the sheet current is an odd function of $x$, we use the integral formulation of the field of section~\ref{sec:appB_antisymm} of appendix B, i.e. (\ref{eq:Hx_appB2}), (\ref{eq:Hy_appB2}).
The mixed boundary conditions defining the problem are then
\begin{align*}
S_{ - 1/4,0} \psi (x) &= (\pi x/2)^{ - 1/2} H_0 & (0<x<b)\\
S_{1/4,0} \psi (x) &= (\pi x/2)^{ - 1/2} {\textstyle{1 \over 2}}J_c & (b<x<a)\\
S_{1/4,0} \psi (x) &= 0 & (a<x<\infty).
\end{align*}
Since the critical point is $b$, we divide the real positive axis in the two intervals $I_1 :(0,b) \cup I_2 :(b,\infty )$, and hence we define $f = f_1 \dot  + f_2$ and $g = g_1 \dot  + g_2$,  where the following parts are known
\begin{align*}
f_1 (x) &= (\pi x/2)^{ - 1/2} H_0 \\
g_2 (x) &= \left\{ {\begin{array}{*{20}c}
   {(\pi x/2)^{ - 1/2} {\textstyle{1 \over 2}}J_c \quad (b < x < a)}  \\
   {0\quad \quad \quad \quad \quad \quad(a<x<\infty)}.  \\
\end{array}} \right.
\end{align*}
By means of these definitions the operator system can be rewritten in a formally identical way to that of the perfect conductor (section~\ref{sec:pc_field})
\begin{align}
\tag{i}
S_{ - 1/4,0} \psi (x) &= f_1 \dot +f_2 \\
\tag{ii}
S_{1/4,0} \psi (x) &= g_1 \dot +g_2.
\end{align}
Since we know the couple $(f_1,g_2)$, we shall use the second mode of table~\ref{tab:tape_field}. 
Making the known parts explicit in the two intervals, we obtain the following equations
\begin{align*}
f_1 (x) &= \left( {\begin{array}{*{20}c}
   x  \\
   0  \\
\end{array}} \right)I_{1/4, - 1/2} \varphi _1 (x) & (x \in I_1)\\
g_2 (x) &= \left( {\begin{array}{*{20}c}
   \infty   \\
   x  \\
\end{array}} \right)K_{1/4, - 1/2} \varphi _2 (x) & (x \in I_2).
\end{align*}
By inversion we can therefore obtain the full function $\varphi  = \varphi _1 \dot  + \varphi _2$  whose parts are
\begin{align*}
\varphi _1 (x) &=  \left( {\begin{array}{*{20}c}
   x  \\
   0  \\
\end{array}} \right)I_{ - 1/4,1/2} f_1 (x) = (\pi /2)^{ - 1/2} H_0 I_{-1/4,1/2} (x^{ - 1/2} ) = \sqrt 2 H_0 x^{ - 1/2} \quad (0 < x < b)
\\
\varphi _2 (x) &= \left( {\begin{array}{*{20}c}
   \infty   \\
   x  \\
\end{array}} \right)K_{1/2, - 1/2} g_2 (x) = \left\{ {\begin{array}{*{20}c}
   {(\sqrt 2 J_c /\pi )x^{ - 1/2} \cosh ^{ - 1} (a/x)\quad (b < x < a)}  \\
   {0\quad \quad (a<x<\infty)}  \\
\end{array}} \right..
\end{align*}
According to (\ref{eq:Jz_appB2}), the current density then will be
\begin{equation*}
J_z (x) =   2(\pi x/2)^{1/2} g(x) = 2(\pi x/2)^{1/2} K_{1/4, - 1/2} \varphi (x).
\end{equation*}
and explicitly in the interval $I_1$
\begin{align*}
J_z (x) &= 2(\pi x/2)^{1/2} \left\{ {\left( {\begin{array}{*{20}c}
   b  \\
   x  \\
\end{array}} \right)K_{\frac{1}{4}, - \frac{1}{2}} \varphi _1 (x) + \left( {\begin{array}{*{20}c}
   a  \\
   b  \\
\end{array}} \right)K_{\frac{1}{4}, - \frac{1}{2}} \varphi _2 (x)} \right\}
\\
&=  - \sqrt 2  \left\{ { \frac{d}{{dx}}[{\bf I}_1 (x) + {\bf I}_2 (x)]} \right\},
\end{align*}
where
\begin{eqnarray*}
{\bf I}_1 (x) &=& \int\limits_x^b {\frac{{u^{3/2} \varphi _1 (u)}}{{\sqrt {u^2  - x^2 } }}du = \sqrt 2 H_0 \sqrt {b^2  - x^2 } } \\
{\bf I}_1 (x) &=& \int\limits_b^a {\frac{{u^{3/2} \varphi _2 (u)}}{{\sqrt {u^2  - x^2 } }}du} = \frac{{\sqrt 2 }}{\pi }J_c \left[ {a\tan ^{ - 1} \sqrt {\frac{{a^2  - b^2 }}{{b^2  - x^2 }}}  - x\tan ^{ - 1} \frac{x}{a}\sqrt {\frac{{a^2  - b^2 }}{{b^2  - x^2 }}}  - \sqrt {b^2  - x^2 } \cosh ^{ - 1} \frac{a}{b}} \right].
\end{eqnarray*}
Carrying out the derivatives, we obtain
\begin{equation*}
\frac{d}{{dx}}\left[ {{\bf I}_1 (x) + {\bf I}_2 (x)} \right] = \frac{x}{{\sqrt {b^2  - x^2 } }}\left( {\frac{{\sqrt 2 }}{\pi }J_c \cosh ^{ - 1} \frac{a}{b} - \sqrt 2 H_0 } \right) - \frac{{\sqrt 2 }}{\pi }J_c \tan ^{ - 1} \frac{x}{a}\sqrt {\frac{{a^2  - b^2 }}{{b^2  - x^2 }}}. 
\end{equation*}
The first term presents a singularity at the critical point $x=b$, which creates a discontinuity of the current density; this is physically unacceptable, since the current density discontinuity is in the interior of the tape. As a consequence, it is necessary to null the part in parentheses
\begin{equation*}
H_0  - \frac{1}{\pi }J_c \cosh ^{ - 1} \frac{a}{b} = 0.
\end{equation*}
From this we can derive the (hitherto arbitrary) value of $b$, which defines the null field zone as a function of the amplitude of the applied field
\begin{equation}\label{eq:b_tapes}
b = a/\cosh (H_0 /H_c ),
\end{equation}
where $H_c=J_c/\pi$. The current density in $I_1$ will be
\begin{equation}
J_z (x) = \frac{2}{\pi }J_c \tan ^{ - 1} \frac{x}{a}\sqrt {\frac{{a^2  - b^2 }}{{b^2  - x^2 }}}. 
\end{equation}
The total transverse magnetic field is given by
\begin{equation}\label{eq:Hy_tapes}
H_y (x) = H_0  - \,(\pi x/2)^{1/2} f(x) = H_0  - (\pi x/2)^{1/2} I_{1/4, - 1/2} \varphi (x),
\end{equation}
and in particular in the band $(b,a)$ where $J=J_c$ the total field will be
\begin{align*}
H_y (x) &= H_0  - (\pi x/2)^{1/2} \left\{ {\left( {\begin{array}{*{20}c}
   b  \\
   0  \\
\end{array}} \right)I_{1/4, - 1/2} \varphi _1 (x) + \left( {\begin{array}{*{20}c}
   x  \\
   b  \\
\end{array}} \right)I_{1/4, - 1/2} \varphi _2 (x)} \right\}
\\
&=   H_0  - \frac{1}{{\sqrt 2 }}\frac{d}{{dx}}[{\bf I}_3 (x) + {\bf I}_4 (x)] \quad \quad (b<x<a),
\end{align*}
where
\begin{eqnarray*}
{\bf I}_3 (x) &=& \int\limits_0^b {\frac{{u^{3/2} }}{{\sqrt {x^2  - u^2 } }}\varphi _1 (u)du = } \sqrt 2 H_0 \left( {x - \sqrt {x^2  - b^2 } } \right) \\
{\bf I}_4 (x) &=& \int\limits_b^x {\frac{{u^{3/2} }}{{\sqrt {x^2  - u^2 } }}\varphi _2 (u)du} \\
&=& \frac{{\sqrt 2 }}{\pi }J_c \left( {\sqrt {x^2  - b^2 } \cosh ^{ - 1} \left( {\frac{a}{b}} \right) - x\coth ^{ - 1} \left( {\frac{x}{a}\sqrt {\frac{{a^2  - b^2 }}{{x^2  - b^2 }}} } \right) + a\ln \frac{{\sqrt {a^2  - b^2 }  + \sqrt {x^2  - b^2 } }}{{\sqrt {a^2  - x^2 } }}} \right).
\end{eqnarray*}
By performing the derivatives we obtain
\begin{align}\label{eq:Hy_tape_sc_magn}
H_y (x) &= \frac{{J_c }}{\pi }\tanh ^{ - 1} \frac{a}{x}\sqrt {\frac{{x^2  - b^2 }}{{a^2  - b^2 }}} \quad \quad (b<x<a).
\end{align}
Outside the tape the magnetic field is still given by~(\ref{eq:Hy_tapes}), but with different integration extremes
\begin{align*}
H_y (x) &= H_0  - (\pi x/2)^{1/2} \left\{ {\left( {\begin{array}{*{20}c}
   b  \\
   0  \\
\end{array}} \right)I_{1/4, - 1/2} \varphi _1 (x) + \left( {\begin{array}{*{20}c}
   a  \\
   b  \\
\end{array}} \right)I_{1/4, - 1/2} \varphi _2 (x)} \right\}\quad (a<x<\infty)
\\
&=  H_0  - \frac{1}{{\sqrt 2 }}\frac{d}{{dx}}[{\bf I}_3 (x) + {\bf I}_5 (x)] \quad \quad (a<x<\infty),
\end{align*}
where
\begin{align*}
{\bf I}_5 (x) = \frac{{\sqrt 2 }}{\pi }J_c \int\limits_b^a {\frac{{u^{} }}{{\sqrt {x^2  - u^2 } }}\cosh ^{ - 1} (a/u)du} \quad \quad (a<x<\infty).
\end{align*}
Solving the integrals, we can easily find
\begin{equation}
H_y (x) = \left( {\frac{{J_c }}{\pi }\cosh ^{ - 1} (a/b) - H_0 } \right)\frac{x}{{\sqrt {x^2  - b^2 } }} - \frac{{J_c }}{\pi }\tanh ^{ - 1} \frac{x}{a}\sqrt {\frac{{a^2  - b^2 }}{{x^2  - b^2 }}} \quad \quad (a<x<\infty).
\end{equation}
Exploiting again the Norris method, we can obtain the ac losses by integrating the magnetic flux in the side bands
\begin{align*}
L_c  = 8\mu _0 J_c \int\limits_b^a {(x - b)H_y (x)dx}. 
\end{align*}
By means of~(\ref{eq:Hy_tape_sc_magn}) this integral can be analytically solved. Introducing a reference loss value $L_0  = 4\mu _0 a^2 J_c H_0  = 4\mu _0 a^2 J_c^2 p/\pi$     where $p=H_0/H_c$,  we immediately recover the formula of Brandt-Indenbom -- see formula (3.26) in~\cite{J:1993:Brandt93}
\begin{equation*}
\frac{{L_c }}{{L_0 }} = \frac{2}{{a^2 p}}\int\limits_b^a {(a - x)\tanh ^{ - 1} \left( {\frac{a}{x}\sqrt {\frac{{x^2  - b^2 }}{{a^2  - b^2 }}} } \right)dx = \frac{2}{p}\ln \cosh p - \tanh p}. 
\end{equation*}

\section{Superconducting plane with a gap}
\begin{figure}[ht!]
\centering
\includegraphics[width=0.45\textwidth] {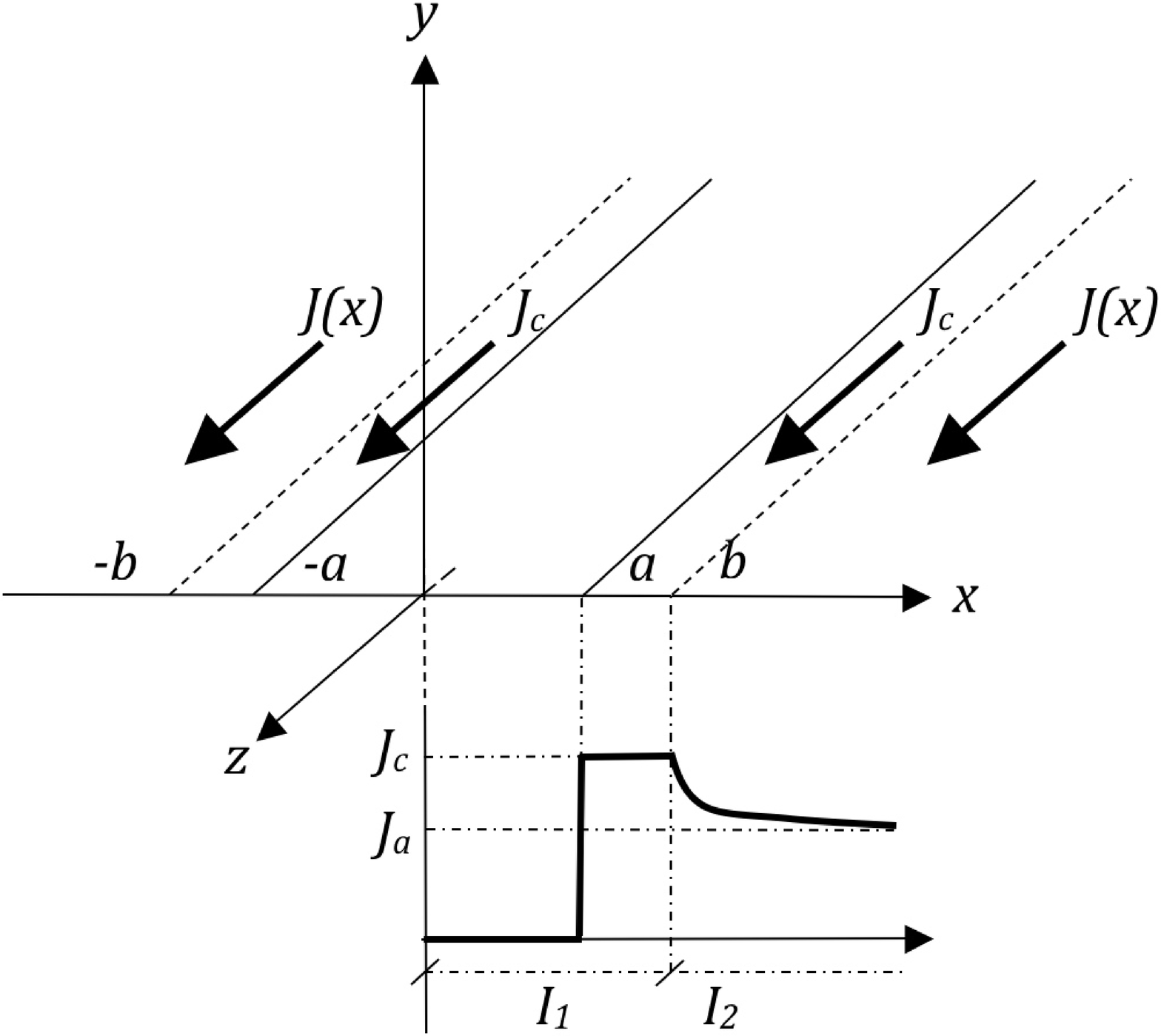}
\caption{Superconducting plan with a gap carrying transport current}
\label{fig:gap_sc}
\end{figure}

We conclude this work discussing a case that requires a little more complex development: an infinite plane ($y=0$) with a transport sheet current $J_a<J_c$ in the $z$-direction with a gap $(-a,a)$ along the $x$-axis, as schematically shown in figure~\ref{fig:gap_sc}. This case has also been treated by Norris~\cite{J:1970:Norris70} and Majoros~\cite{J:1996:Majoros96}. On the planeÕs edges near the gap there will be two regions $(-b,-a)$ and $(a,b)$ where the sheet current assumes the critical value $J_c$ and where the magnetic flux penetrates, causing dissipation. Our goal is to find the width of these two regions, i.e. the point $b$, and to compute the value of the transversal magnetic field $H_y$, in order to be able to derive the magnetic flux and, from it, the value of the ac losses.

\begin{figure}[h!]
\centering
\includegraphics[width=0.45\textwidth] {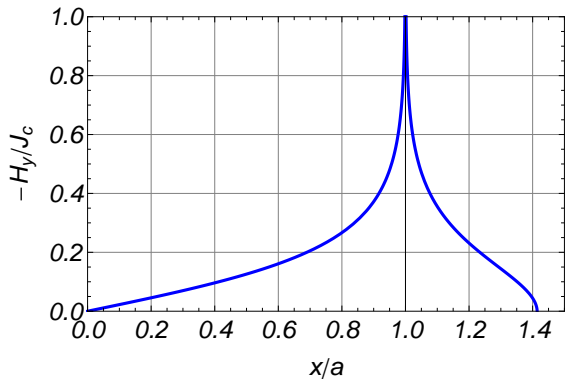}
\caption{Magnetic field (normalized to the sheet current $J_c$) profile in the interval $(0,b)$ for a sheet current $J_a=0.5J_c$.}
\label{fig:H_gap}
\end{figure}

The problem presents an obvious even symmetry and it will be sufficient to consider the region with $x>0$, which can be divided in three intervals: (1) the gap $(0,a)$, where $H_x=0$ by symmetry; (2) the interval $(a,b)$ where $J=J_c$ and therefore $H_x=-J_c/2$; (3) the interval $(b,\infty)$ where $H_y=0$. Since in (1) and (2) there is the same type of boundary conditions, the two intervals can be merged into one, so that the whole domain is finally divided by the critical point $b$ in the two intervals $I_1 :(0,b)$  and $I_2 :(b,\infty)$, where we have the following mixed boundary conditions
\begin{align*}
H_x  &= \left\{ {\begin{array}{*{20}c}
   {0\quad (0 < x < a)}  \\
   {-{\textstyle{1 \over 2}}J_c \quad (a < x < b)} 
\end{array}} \right. & (x \in I_1)\\
H_y  &= 0 & (x \in I_2).
\end{align*}
By defining as usual $f=f_1 \dot + f2$ and $g=g_1 \dot + g_2$, where 
\begin{align*}
& g_1  = (\pi x/2)^{ - 1/2} \left\{ {\begin{array}{*{20}c}
   {0\quad \quad (0 < x < a)}  \\
   {{\textstyle{1 \over 2}}J_c \quad \quad (a < x < b)}  \\
\end{array}} \right. \\
& f_2=0,
\end{align*}
and by using expressions (\ref{eq:Hx_appB}-\ref{eq:Hy_appB}) of appendix B, the previous conditions can be written as
\begin{align}
\tag{i}
S_{ - 1/4,0} \psi  = g_1 \dot+ g_2 \\
\tag{ii}
S_{1/4,0} \psi  = f_1 \dot+ f_2.
\end{align}
Let us try the third mode of table~\ref{tab:tape_curr}. From the first equation developed in $I_2$, since $f_2=0$, we derive the following equation
\begin{align*}
0 = \left( {\begin{array}{*{20}c}
   b  \\
   0  \\
\end{array}} \right)I_{ - 1/4,1/2} g_1  + \left( {\begin{array}{*{20}c}
   x  \\
   b  \\
\end{array}} \right)I_{ - 1/4,1/2} g_2 
\quad \quad (x \in I_2)
\end{align*}
and inverting it we have
\begin{align}
\nonumber
g_2  &=  - \left( {\begin{array}{*{20}c}
   x  \\
   b  \\
\end{array}} \right)I_{ - 1/4,1/2}^{ - 1} \left( {\begin{array}{*{20}c}
   b  \\
   0  \\
\end{array}} \right)I_{ - 1/4,1/2} g_1  =  - \left( {\begin{array}{*{20}c}
   x & b  \\
   b & 0  \\
\end{array}} \right)L_{ - 1/4,1/2} g_1 \\
\nonumber
&=  
- \frac{2}{\pi }({\textstyle{1 \over 2}}J_c )(\pi /2)^{ - 1/2} \frac{{x^{1/2} }}{{\sqrt {x^2  - b^2 } }}\int\limits_a^b {\frac{{\sqrt {b^2  - t^2 } }}{{x^2  - t^2 }}dt} \\
\label{eq:f2_gap}
&= (\pi /2)^{ - 3/2} ({\textstyle{1 \over 2}}J_c )\frac{{x^{1/2} }}{{\sqrt {x^2  - b^2 } }}\left( {\frac{{\sqrt {x^2  - b^2 } }}{x}\tan ^{ - 1} \frac{x}{a}\sqrt {\frac{{b^2  - a^2 }}{{x^2  - b^2 }}}  - \cos ^{ - 1} \frac{a}{b}} \right).
\end{align}
Knowing the two parts of $g$,  we can compute the sheet current density. From equation~(\ref{eq:Jz_appB}) of appendix B we have $J = 2(\pi x/2)^{1/2} g$  and therefore in $I_2$ we shall have
\begin{align}
J(x) = \frac{2}{\pi }J_c \left( {\tan ^{ - 1} \frac{x}{a}\sqrt {\frac{{b^2  - a^2 }}{{x^2  - b^2 }}}  - \frac{x}{{\sqrt {x^2  - b^2 } }}\cos ^{ - 1} \frac{a}{b}} \right) \quad \quad (x \in I_2).
\end{align}
Unfortunately, this expression for the sheet current has two flaws:
\begin{itemize}
\item in $x=b$ it goes to infinity, which is not physically acceptable;
\item since $\mathop {\lim }\limits_{x \to \infty } \tan ^{ - 1} \frac{x}{a}\sqrt {\frac{{b^2  - a^2 }}{{x^2  - b^2 }}}  = \cos ^{ - 1} \frac{a}{b}$, the sheet current $J(x)$ approaches 0 instead of  the asymptotic value $J_a$ requested by the problem.
\end{itemize}
In order to overcome these problems, we choose a more general $H_x$ in $I_1$
\begin{align*}
H_x  = \left\{ {\begin{array}{*{20}c}
   { - c_1 \quad (0 < x < a)}  \\
   { - c_2 \quad (a < x < b)}  \\
\end{array}} ,\right.
\end{align*}
where the constants $c_1$ and $c_2$ will be determined later, in order to satisfy the conditions imposed by the problem. Consequently, we have
\begin{align*}
g_1 (x) = (\pi x/2)^{ - 1/2} \left\{ {\begin{array}{*{20}c}
   {c_1 \quad \quad (0 < x < a)}  \\
   {c_2 \quad \quad (a < x < b)}  \\
\end{array}} \right. 
\end{align*}
and inserting it in equation~(\ref{eq:f2_gap}) we obtain
\begin{align*}
g_2 (x) &=  - \frac{2}{\pi }\frac{{x^{1/2} }}{{\sqrt {x^2  - b^2 } }}\int\limits_0^b {\frac{{\sqrt {b^2  - t^2 } }}{{x^2  - t^2 }}t^{1/2} g_1 (t)dt} \\
&= - \frac{2}{\pi }(\pi x/2)^{ - 1/2} \frac{x}{{\sqrt {x^2  - b^2 } }}
\left[ {c_1 {\bf I}_1  + c_2 {\bf I}_2 } \right]
\end{align*}%
with
\begin{align*}
{\bf I}_1 (x) &= \int\limits_0^a {\frac{{\sqrt {b^2  - t^2 } }}{{x^2  - t^2 }}dt}  = \frac{\pi }{2} - \cos ^{ - 1} \frac{a}{b} - \frac{{\sqrt {x^2  - b^2 } }}{x}\left( {\frac{\pi }{2} - \tan ^{ - 1} \frac{{x\sqrt {b^2  - a^2 } }}{{a\sqrt {x^2  - b^2 } }}} \right) \\
{\bf I}_2 (x) &= \int\limits_a^b {\frac{{\sqrt {b^2  - t^2 } }}{{x^2  - t^2 }}dt}  = \cos ^{ - 1} (a/b) - \tan ^{ - 1} \left( {\frac{x}{a}\sqrt {\frac{{b^2  - a^2 }}{{x^2  - b^2 }}} } \right).
\end{align*}
The sheet current can be written as a two-part function $J_ (x) = J_1(x)\dot+ J_2 (x)$  where
\begin{align*}
& J_1 (x) = 2(\pi x/2)^{1/2} g_1  = \left\{ {\begin{array}{*{20}c}
   {2c_1 \quad \quad (0 < x < a)}  \\
   {2c_2 \quad \quad (a < x < b)}  \\
\end{array}} \right. & \quad \quad(x \in I_1)\\
& J_2 (x) = 2(\pi x/2)^{1/2} g_2  =  - \frac{4}{\pi }\frac{x}{{\sqrt {x^2  - b^2 } }}\left[ {c_1 {\bf I}_1  + c_2 {\bf I}_2 } \right] & \quad \quad(x \in I_2),
\end{align*}
which, by using the analytical expressions of the integral, becomes
\begin{align*}
 J_2 (x) &=  - \frac{{2x}}{{\sqrt {x^2  - b^2 } }}\left[ {c_1  + \frac{2}{\pi }(c_2  - c_1 )\cos ^{ - 1} (a/b)} \right] \\ 
&+ 2c_1  + (c_2  - c_1 )\frac{4}{\pi }\tan ^{ - 1} \frac{x}{a}\sqrt {\frac{{b^2  - a^2 }}{{x^2  - b^2 }}}.
\end{align*}
With this new approach to the problem we can now remove the divergence to infinity at the critical  $x=b$ if we null the expression in brackets

\begin{align*}
J_1  + \frac{2}{\pi }(J_2  - J_1 )\cos ^{ - 1} (a/b) = 0.
\end{align*}
From this expression we determine the value of $b$
\begin{align}\label{eq:b_gap}
b = \frac{a}{{\cos \left( {\frac{\pi }{2}\frac{{c_1 }}{{c_2  - c_1 }}} \right)}},
\end{align}
the beginning of the null field zone. The expression for the current density reduces to
\begin{align*}
J_2 (x) = 2c_1  + (c_2  - c_1 )\frac{4}{\pi }\tan ^{ - 1} \frac{x}{a}\sqrt {\frac{{b^2  - a^2 }}{{x^2  - b^2 }}} . 
\end{align*}
In order to eliminate the component $J_1(x)$  in the gap $0<x<a$, we need to add a current layer $-2c_1$ in the whole plane. This added current does not change the profile of $H_y$ nor the position of $b$. We will have
\begin{align}\label{eq:J1_gap}
J_1 (x) &= \left\{ {\begin{array}{*{20}c}
   {0\quad \quad (0 < x < a)}  \\
   {2(c_2  - c_1 )\quad \quad (a < x < b)}  \\
\end{array}} \right. \\
\label{eq:J2_gap}
J_2 (x) &= (c_2  - c_1 )\frac{4}{\pi }\tan ^{ - 1} \frac{x}{a}\sqrt {\frac{{b^2  - a^2 }}{{x^2  - b^2 }}}. 
\end{align}
In order to determine the (hitherto arbitrary) constants $c_1$ and $c_2$, we impose the conditions
\begin{align*}
J_2 (b) = J_c  = 2(c_2  - c_1 ),
\end{align*}
from which $c_2  - c_1  = J_c /2$. Taking into account the limit as $x$ approaches infinity, we shall also have 
\begin{align*}
J_2 (\infty ) = J_a  = (c_2  - c_1 )\frac{4}{\pi }\cos ^{ - 1} \frac{a}{b}
\end{align*}
and, by equation~(\ref{eq:b_gap}), $\cos ^{ - 1} a/b = \frac{\pi }{2}\frac{{c_1 }}{{c_2  - c_1 }}$, so that $2c_1=J_a$ from which $c_1=J_a/2$.
Inserting the expressions for $c_1$ and $c_2$ in~(\ref{eq:J1_gap}) and ~(\ref{eq:J2_gap}), we obtain the current density distribution
\begin{align}
\nonumber
& J_1 (x) = \left\{ {\begin{array}{*{20}c}
   {0\quad \quad (0 < x < a)}  \\
   {J_c \quad \quad (a < x < b)}  \\
\end{array}} \right. & (x \in I_1) \\
\nonumber
& J_2 (x) = J_c \frac{2}{\pi }\tan ^{ - 1} \frac{x}{a}\sqrt {\frac{{b^2  - a^2 }}{{x^2  - b^2 }}} & (x \in I_2) \\
& b = a/\cos \left( {\frac{\pi }{2}\frac{{J_a }}{{J_c }}} \right). &
\end{align}
\begin{figure}[h!]
\centering
\includegraphics[width=0.45\textwidth] {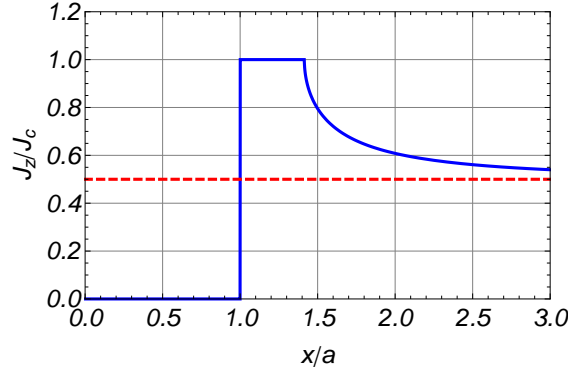}
\caption{Sheet current distribution in an infinite plane carrying a sheet current $J_a=0.5J_c$ (represented by the dashed line) with a gap in $(-a,a)$. Only the semi-plane $x>0$ is shown. Near the gap the sheet current is equal to $J_c$ over a distance $b$ (in this case $b = \sqrt{2}a$), then it decays toward the asymptotic value $J_a$. See also figure~\ref{fig:gap_sc} for reference.}
\label{fig:J_gap}
\end{figure}
An example of the current density profile is given in figure~\ref{fig:J_gap}. Once the current density is known in the whole plane, it is possible to compute the transversal magnetic field in the interval $I_1$ (in the interval $I_2$ the magnetic field is null) by means of the Biot-Savart formula applied to the whole $x$-axis
\begin{equation*}
H_y (x) = \frac{1}{{2\pi }}\int\limits_0^\infty  {\frac{{J_z (\xi )}}{{x - \xi }}} d\xi  + \frac{1}{{2\pi }}\int\limits_0^\infty  {\frac{{J_z (\xi )}}{{x + \xi }}} d\xi, 
\end{equation*}
where the first integral must obviously be calculated as principal value. However this can be avoided by using formulas (\ref{eq:Hy_appB}) and~(\ref{eq:Jz_appB}) of appendix B, i.e.  
\begin{align*}
& H_y  = (\pi x/2)^{1/2} S_{1/4,0} \psi 
 \\
& J_z  =  - 2H_x  = 2(\pi x/2)^{1/2} S_{ - 1/4,0} \psi.
\end{align*}
Eliminating $\psi$, we can write
\begin{equation}\label{eq:Hy_gap0}
H_y (x) = \frac{{1}}{2}x^{1/2} (S_{1/4,0} S_{ - 1/4,0} )x^{ - 1/2} (J_1 \dot  + J_2 ).
\end{equation}
Since there is no combination rule for the two $S$ operators, expression~(\ref{eq:Hy_gap0}) has to be explicitly computed. We can write the magnetic field as the sum of the two magnetic fields generated by the currents in the two intervals: $H_y (x) = H' (x) + H'' (x)$, where
\begin{align*}
H' (x) &= \frac{{J_c }}{2}x^{1/2} (S_{1/4,0} S_{ - 1/4,0} )[x^{ - 1/2} ] = \frac{{J_c }}{2}x^{1/2} S_{1/4,0} \left\{ {\int\limits_a^b {t^{1/2} J_{ - 1/2} (xt)dt} } \right\} \\
&= \frac{{J_c }}{2}x^{1/2} S_{1/4,0} \left\{ {\sqrt {\frac{2}{\pi }} \frac{{\sin bt - \sin at}}{{x^{3/2} }}} \right\} \\
&= \sqrt {\frac{2}{\pi }} \frac{{J_c }}{2}x^{1/2} \int\limits_0^\infty  {t^{ - 1/2} } (\sin bt - \sin at)J_{1/2} (xt)dt \\
&= \frac{{J_c }}{\pi }\left( {\tanh ^{ - 1} x/b - \tanh ^{ - 1} a/x} \right) \\
H'' (x) &= \frac{{J_c }}{\pi }x^{1/2} (S_{1/4,0} S_{ - 1/4,0} )[x^{ - 1/2} \tan ^{ - 1} \left( {\frac{x}{a}\sqrt {\frac{{b^2  - a^2 }}{{x^2  - b^2 }}} } \right)] \\
&= \frac{{J_c }}{\pi }x^{1/2} S_{1/4,0} \chi (x),
\end{align*}
with
\begin{align*}
\chi (x) = \int\limits_b^\infty  {t^{1/2} } \tan ^{ - 1} \left( {\frac{t}{a}\sqrt {\frac{{b^2  - a^2 }}{{t^2  - b^2 }}} } \right)J_{ - 1/2} (xt)dt.
\end{align*}
Changing the order of integration, we have
\begin{align*}
S_{1/4,0} \chi (x) &= \int\limits_0^\infty  {u\,J_{1/2} (xu)du\int\limits_b^\infty  {t^{1/2} \tan ^{ - 1} 
\left( {\frac{t}{a}\sqrt {\frac{{b^2  - a^2 }}{{t^2  - b^2 }}} } \right)
J_{ - 1/2} (ut)dt} } \\
&= \int\limits_b^\infty  {t^{1/2} \tan ^{ - 1} 
\left( {\frac{t}{a}\sqrt {\frac{{b^2  - a^2 }}{{t^2  - b^2 }}} } \right)
dt} \int\limits_0^\infty  {u\,J_{1/2} (xu)J_{ - 1/2} (tu)} \,du.
\end{align*}
The second integral is a Weber-Shafheitlin integral and in the case $x<t$ one has
\begin{equation*}
S_{1/4,0} \chi (x) = \int\limits_b^\infty  {t^{1/2} \tan ^{ - 1} 
\left( {\frac{t}{a}\sqrt {\frac{{b^2  - a^2 }}{{t^2  - b^2 }}} } \right)
} \left( { - \frac{2}{\pi }\sqrt {\frac{x}{t}} \frac{1}{{t^2  - x^2 }}} \right)dt =  - \frac{2}{\pi }x^{1/2} \int\limits_b^\infty  {\frac{{\tan ^{ - 1} 
\left( {\frac{t}{a}\sqrt {\frac{{b^2  - a^2 }}{{t^2  - b^2 }}} } \right)
}}{{t^2  - x^2 }}} 
\end{equation*}
and
\begin{align*}
H''(x) &= \frac{{J_c }}{\pi }\frac{{2x}}{\pi }\int\limits_b^\infty  {\tan ^{ - 1} \left( {\frac{t}{a}\sqrt {\frac{{b^2  - a^2 }}{{t^2  - b^2 }}} } \right)} \frac{{dt}}{{x^2  - t^2 }} \\
&= \frac{{J_c }}{\pi }\left( {\tanh ^{ - 1} x/b + \tanh ^{ - 1} a/x - \tanh ^{ - 1} \left( {\frac{a}{x}\sqrt {\frac{{b^2  - x^2 }}{{b^2  - a^2 }}} } \right)} \right),
\end{align*}
so that the total magnetic field will be
\begin{align}\label{eq:Hy_gap}
H_y (x) =  - \frac{{J_c }}{\pi }\tanh ^{ - 1} \left( {\frac{a}{x}\sqrt {\frac{{b^2  - x^2 }}{{b^2  - a^2 }}} } \right) \quad \quad (x \in I_1).
\end{align}
The profile of such field in the interval $I_1$ is plotted in figure~\ref{fig:H_gap}.
In order to compute the ac losses we can use Norris's method linking the losses and the magnetic flux in the region $(a,b)$ at the peak of the current 
\begin{align*}
L_c  = 8\mu _0 J_c \int\limits_a^b {(\xi  - a)} H_y (\xi )d\xi.
\end{align*}
By means of (\ref{eq:Hy_gap}), this integral can be analytically computed. Using the normalizing factor introduced by Norris, 
$L_0  = \mu _0 (2aJ_c )^2 /\pi $, we obtain the same formula that can be found in~\cite{J:1996:Majoros96}
\begin{equation*}
L_c /L_0  = 2\ln \left| {\cos (p\pi /2)} \right| + (p\pi /2)\tan (p\pi /2),
\end{equation*}
where $p=J_a/J_c$.
%
\section{Concluding remarks} 
In this paper our foremost intention was to demonstrate the effectiveness of modeling thin superconductors as mixed boundary value problems in those cases where they can be expressed, with the help of appropriate Green functions, in an operator form by means of Hankel and Erd\'elyi-Kober operators. We showed that, thanks to the numerous disentangling properties of these operators, for many problems with a two part mixed boundary subdivision, the resulting equations are reduced to the computation of simple integrals which do not require the principal value interpretation.  Each problem, as shown here, can be reduced to three equivalent modes, each involving different combinations of the operators in order to find the `missing part' of the functions expressing the mixed boundary condition. In general, one mode can be more workable than another, since it produces simple integrals or integral equations that can be solved analytically or by standard numerical method; as a consequence, the choice of the correct mode is the most critical step towards a simple solution. An initial test of all the three modes can reserve the most useful hints since the `wrong' mode reveals itself very quickly by the presence of operator combinations that cannot be simplified by the properties shown in appendix A. 

In all the cases presented here, the mixed boundaries are composed of two intervals of the positive real axis, where $H_x$ or $H_y$  components are known respectively. In more complex cases (always exhibiting some symmetry) it may happen that a subdivision in more intervals is needed. For instance, in the case of a ring in external field or of two adjacent parallel tapes, the real axis is divided in three parts where different components of the field are known; then, although the operator equations are essentially the same as those presented here, their development on each interval has three terms: as a consequence, the extraction of the unknown parts requires multiple combinations of the EK operators with a subsequent complication of the analytical/numerical activity.
\footnote
{For example, in the case of a perfectly conducting ring of radii $b$ and $a$ ($b<a$) subjected to a uniform magnetic field  $-H_0$  we divide the positive axis in the three intervals  $I_1:(0,b)$, $I_2:(b,a)$, $I_3:(a,\infty)$, where the induced currents will generate the mixed boundary conditions $H_r=0$, $H_z=H_0$, $H_r=0$ in $I_1$, $I_2$, $I_3$, respectively.
Using the Green functions~(\ref{eq:appB_circ1}-\ref{eq:appB_circ2}) of appendix~\ref{appB_circ} we can write
\begin{align*}
& S_{1/2,0} \psi  = g_1 \dot  + g_2 \dot  + g_3
\\
& S_{0,0} \psi  = f_1 \dot  +  f_2 \dot  + f_3 
\end{align*}
where $g_1=g_3=0$ and $f_2=H_0$. Using the first mode of table~\ref{tab:disc_unif} we obtain the operator system $f = K_{0, 1/2} \varphi$, $g = I_{0,  1/2} \varphi$. Having developed the system on the intervals, it is then straightforward to obtain the integral equation for $\varphi_2$ 
\begin{equation*}
\varphi _2 (r) = \frac{{H_0 }}{{\sqrt \pi  }}\frac{r}{{\sqrt {a^2  - r^2 } }} + \frac{1}{{\sqrt {a^2  - r^2 } }}\int\limits_b^a {K(x,t)\sqrt {a^2  - t^2 } } \varphi _2 (t)dt
\end{equation*}
with kernel
\begin{equation*}
K(r,t) = \frac{2}{{\pi ^2 (r^2  - t^2 )}}\left( {r\ln \frac{{a - t}}{{a + t}} - t\ln \frac{{a - r}}{{a + r}}} \right).
\end{equation*}
Even though the analytical solution appears to be prohibitive, the numerical solution can be easily found by Nystr\"om method since the equation is not singular. The case of a superconducting ring is also solvable by a similar equation, but the removal of infinities at the critical points that fall inside the ring requires much more efforts. 
We shall present a selection of these difficult cases in a future paper. 
}

The most outstanding feature of the operator method is that it provides a unified and robust rationale for affording large amounts of cases which do not require any {\it ad hoc} methods nor any proviso for each of them, as for example the conformal transform and images. Also it does not compel resorting to integral equation of Cauchy type so that all the computations can be kept in the domain of real analysis, a very welcome quality in the case of numerical work.

We encourage the interested reader to try to solve other similar cases utilizing the operators and their properties we presented here in appendix A and B, in conjunction with the three mode reduction of section 2. He/she will be quickly comforted by the simplicity and immediateness of the procedure, which develops without any ambiguity or necessity of external aids other than the numerical skills required in handling singular integrals. Common software packages for symbolic calculation (such as Mathematica or Maple) are of great support for that task. Analogous electrostatics cases can be handled exactly in the same way once the Green functions for charge lines and circles are provided to play the same role of those given in appendix B.

\section*{Acknowledgments}
This work has been supported partly by the Helmholtz-University Young Investigator Grant VH-NG-617 and partly by the Research Fund for the Italian Electrical System under the Contract Agreement between RSE and the Ministry of Economic Development - General Directorate for Nuclear Energy, Renewable Energy and Energy Efficiency stipulated on July 29, 2009 in compliance with the Decree of March 19, 2009.

\newpage
\appendix
\section{Erd\'elyi-Kober and Hankel operators: a short review}\label{app_A}
\subsection{Erd\'elyi-Kober operators}
The Erd\'elyi-Kober operators are defined as the following fractional integrals (the integration interval is shown in the parenthesis in front of them)
\begin{align}
\nonumber
\left( {\begin{array}{*{20}c}
   x  \\
   a  \\
\end{array}} \right)I_{\eta ,\alpha } f(x) &= \frac{{2x^{ - 2(\alpha  + \eta )} }}{{\Gamma (\alpha )}}\int_a^x {(x^2  - t^2 )^{\alpha  - 1} t^{2\eta  + 1} } f(t)dt & (\alpha>0) \\
\label{eq:xa_appA}
 &= \frac{{x^{ - 2(\alpha  + \eta ) - 1} }}{{\Gamma (1 + \alpha )}}\frac{d}{{dx}}\int_a^x {(x^2  - t^2 )^\alpha  t^{2\eta  + 1} } f(t)dt & (-1<\alpha<0) \\
 \nonumber
\left( {\begin{array}{*{20}c}
   b  \\
   x  \\
\end{array}} \right)K_{\eta ,\alpha } f(x) &= \frac{{2x^{2\eta } }}{{\Gamma (\alpha )}}\int_x^b {(t^2  - x^2 )^{\alpha  - 1} t^{1 - 2(\alpha  + \eta )} f(t)dt} & (\alpha>0) \\
\label{eq:bx_appA}
 &=  - \frac{{x^{2\eta  - 1} }}{{\Gamma (1 + \alpha )}}\frac{d}{{dx}}\int_x^b {(t^2  - x^2 )^\alpha  t^{1 - 2(\alpha  + \eta )} } f(t)dt & (-1<\alpha<0).
\end{align}
One should note that the operators with $\alpha <0$ also require a derivation out of the integral. For both kinds of operators the cases with $\alpha=0$  correspond  to the identity
\begin{equation*}
I_{\eta ,0} f(x) = f(x),\quad \quad K_{\eta ,0} f(x) = f(x).
\end{equation*}
It is very important to observe that the operators $I_{\eta ,\alpha }$  still exist (in real analysis) when the integration limits are both fixed and $x$ is greater than the upper limit
\begin{equation*}
\left( {\begin{array}{*{20}c}
   b  \\
   a  \\
\end{array}} \right)I_{\eta ,\alpha } f(x)\quad (x > b)
\end{equation*}
and likewise the operators $K_{\eta ,\alpha }$  exist when the integration limits are both fixed and $x$ is smaller than the lower limit
\begin{equation*}
\left( {\begin{array}{*{20}c}
   b  \\
   a  \\
\end{array}} \right)K_{\eta ,\alpha } f(x)\quad (x < a).
\end{equation*}
\subsection{Modified operator of the Hankel transform}
One defines the modified Hankel operator $S_{\eta ,\alpha } $ (very similar to the Hankel transform) as
\begin{equation}
S_{\eta ,\alpha } f(x) = \left( {\frac{2}{x}} \right)^\alpha  \int_0^\infty  {t^{1 - \alpha } J_{2\eta  + \alpha } (xt)f(t)dt}, 
\end{equation}
so that the usual Hankel transform becomes
\begin{align*}
\int_0^\infty  {x^{p} f(x)J_\nu  (xy)dx}  = \left( {\frac{y}{2}} \right)^{1-p}  S_{\frac{1}{2}{(\nu  + p-1)}, 1-p } f(y).
\end{align*}
By means of the links between sine and cosine and Bessel functions of half order, i.e. $\sin (z) = {(z\pi /2)}^{1/2} J_{1/2} (z)$ and  $\cos (z) = {(z\pi /2)}^{1/2} J_{ - 1/2} (z)$, also sine and cosine transform can be expressed by
$S_{\alpha ,\beta }$ operators. By defining $\psi (\lambda ) = \lambda ^{ - 1/2} F(\lambda )$, we have
\begin{align*}
\int\limits_0^\infty  {F(\lambda )\sin (x\lambda )d\lambda }  = {(\pi x /2)}^{1/2} S_{1/4,0} \psi (x) \\
\int\limits_0^\infty  {F(\lambda )\cos (x\lambda )d\lambda }  = {(\pi x /2)}^{1/2} S_{ - 1/4,0} \psi (x).
\end{align*}

\subsection{Operator inversion}
The inverse operators (which express the solution of Abel's generalized equation) are given by the following expressions
\begin{align}
\label{eq:inv_xa}
\left( {\begin{array}{*{20}c}
   x  \\
   a  \\
\end{array}} \right)I_{\eta ,\alpha }^{ - 1} f(x) = \left( {\begin{array}{*{20}c}
   x  \\
   a  \\
\end{array}} \right)I_{\eta  + \alpha , - \alpha }^{} f(x) \\
\label{eq:inv_bx}
\left( {\begin{array}{*{20}c}
   b  \\
   x  \\
\end{array}} \right)K_{\eta ,\alpha }^{ - 1} f(x) = \left( {\begin{array}{*{20}c}
   b  \\
   x  \\
\end{array}} \right)K_{\eta  + \alpha , - \alpha }^{} f(x). 
\end{align}
As a consequence, operator equations like
\begin{align*}
\left( {\begin{array}{*{20}c}
   x  \\
   a  \\
\end{array}} \right)I_{\eta ,\alpha }^{} f(x) = g(x) \\
\left( {\begin{array}{*{20}c}
   b  \\
   x  \\
\end{array}} \right)K_{\eta ,\alpha }^{} f(x) = g(x)
\end{align*}
can be immediately solved by using~(\ref{eq:xa_appA}) and (\ref{eq:bx_appA})
\begin{align*}
f(x) = \left( {\begin{array}{*{20}c}
   x  \\
   a  \\
\end{array}} \right)I_{\eta ,\alpha }^{ - 1} g(x) = \left( {\begin{array}{*{20}c}
   x  \\
   a  \\
\end{array}} \right)I_{\eta  + \alpha , - \alpha }^{} g(x) \\
f(x) = \left( {\begin{array}{*{20}c}
   b  \\
   x  \\
\end{array}} \right)K_{\eta ,\alpha }^{ - 1} g(x) = \left( {\begin{array}{*{20}c}
   b  \\
   x  \\
\end{array}} \right)K_{\eta  + \alpha , - \alpha }^{} g(x).
\end{align*}
One remarks that the inverse operators have the same integration limits. Also in the case of inversion of EK operators with both fixed integration limits we formally shall write (\ref{eq:inv_xa}) and (\ref{eq:inv_bx}), but unfortunately in these cases general explicit expressions for them are not available. This is because the EK operators no longer represent Abel integral equations of Volterra type, i.e. integral equations with one varying integration limit. Due to their definition also Hankel operators $S_{\eta ,\alpha }$  satisfy an identical inversion rule 

\begin{equation}\label{eq:Setaalpha_appA}
S_{\eta ,\alpha }^{ - 1} f(x) = S_{\eta  + \alpha , - \alpha }^{} f(x)
\end{equation}
as long as the integration interval is the whole positive real axis. 
\subsection{Integral equations fractioned in two parts}
Let us suppose that the real positive axis $x>0$ is divided in two complementary intervals $I_1  = (0 < x < a)$ and $I_2  = (a < x < \infty )$. We shall express a generic function defined on the positive axis  $x>0$ as sum of two parts (which we shall mark by a dotted plus, $f(x) = f_1 (x)\dot  + f_2 (x)$) defined only in the sub-interval of their own index and undefined elsewhere
\begin{align*}
f_i (x) = \left\{ {\begin{array}{*{20}c}
   {f(x) \quad \quad \quad \quad x \in I_i }  \\
   {{\rm undefined}\quad \quad x \notin I_i}.  \\
\end{array}} \right.
\end{align*}
Based on the definition $\varphi(x)=\varphi_1(x) \dot + \varphi_2(x)$, the operator equation 
\begin{equation*}
f(x) = \left( {\begin{array}{*{20}c}
   x  \\
   0  \\
\end{array}} \right)I_{\eta ,\alpha } \varphi (x)
\end{equation*}
splits into the following two equations, one for each interval
\begin{align}
\label{eq:f1_appA}
 f_1 (x) &=  \left( {\begin{array}{*{20}c}
   x  \\
   0  \\
\end{array}} \right)I_{\eta ,\alpha }\varphi _1 (x) & (x \in I_1 ) \\
\label{eq:f2_appA}
 f_2 (x) &= \left( {\begin{array}{*{20}c}
   a  \\
   0  \\
\end{array}} \right)I_{\eta ,\alpha } \varphi _1 (x) +  \left( {\begin{array}{*{20}c}
   x  \\
   a  \\
\end{array}} \right)I_{\eta ,\alpha } \varphi _2 (x) & (x \in I_2 )
\end{align}
and the operator equation
\begin{equation*}
g(x) = \left( {\begin{array}{*{20}c}
   \infty   \\
   x  \\
\end{array}} \right)K_{ \eta , \alpha } \varphi (x)
\end{equation*}
similarly splits into the two following equations:
\begin{align}
\label{eq:g1_appA}
g_1 (x) &= \left( {\begin{array}{*{20}c}
   a  \\
   x  \\
\end{array}} \right)K_{\eta ,\alpha } \varphi _1 (x) + \left( {\begin{array}{*{20}c}
   \infty   \\
   a  \\
\end{array}} \right)K_{\eta ,\alpha } \varphi _2 (x) &(x \in I_1 ) \\
\label{eq:g2_appA}
g_2 (x) &= \left( {\begin{array}{*{20}c}
   \infty   \\
   x  \\
\end{array}} \right)K_{ \eta , \alpha } \varphi _2 (x) & (x \in I_2 ).
\end{align}
The generalization to more than two intervals is evident: we shall have an equation for each subinterval where there is the sum of applications of   $I_{\eta,\alpha}$ to the previous intervals or the sum of applications of   $K_{\eta,\alpha}$ to the remaining intervals.
All but one will have fixed limits of integration so that only one inversion is feasible for each equation by means of~(\ref{eq:inv_xa}) or (\ref{eq:inv_bx}). In the present case, for instance, if we know the couple $(f_1 ,g_2 )$  we immediately obtain $\varphi_1$  from (\ref{eq:f1_appA}) and $\varphi_2$ from (\ref{eq:g2_appA}) . If we know the couple $(f_2 ,g_1 )$  by inversion of (\ref{eq:f2_appA}) and (\ref{eq:g1_appA}) and eliminating $\varphi_1$  or $\varphi_2$  we can obtain an integral equation of Fredholm type of second kind for the remaining part.

\subsection{Double operators}\label{subsec:double_operators}
In the solution process of the previous systems, the following combinations of two operators (fully developed in~\cite{J:1963:Cooke63}) often appear
\begin{align}
\nonumber
&
\left( {\begin{array}{*{20}c}
   x  \\
   a  \\
\end{array}} \right)I_{\eta ,\alpha }^{ - 1} \left( {\begin{array}{*{20}c}
   f  \\
   e  \\
\end{array}} \right)I_{\eta ,\alpha }^{} f
= \left( {\begin{array}{*{20}c}
   x & f  \\
   a & e  \\
\end{array}} \right)L_{\eta ,\alpha } f \\
\label{eq:double_Ina}
&=
\frac{{2\sin \alpha \pi }}{\pi }\frac{{x^{ - 2\eta } }}{{(x^2  - a^2 )^\alpha  }}\int\limits_e^f {\frac{{u^{2\eta  + 1} (a^2  - u^2 )^\alpha  }}{{x^2  - u^2 }}} f(u)du 
\quad \quad (x>a>f>e) \\
\nonumber
&
\left( {\begin{array}{*{20}c}
   b  \\
   x  \\
\end{array}} \right)K_{\eta ,\alpha }^{ - 1} \left( {\begin{array}{*{20}c}
   f  \\
   e  \\
\end{array}} \right)K_{\eta ,\alpha }^{} f 
= \left( {\begin{array}{*{20}c}
   b & f  \\
   x & e  \\
\end{array}} \right)M_{\eta ,\alpha } f \\
\label{eq:double_Kna}
&= 
\frac{{2\sin \alpha \pi }}{\pi }\frac{{x^{2(\alpha  + \eta )} }}{{(b^2  - x^2 )^\alpha  }}\int\limits_e^f {\frac{{u^{1 - 2(\alpha  + \eta )} (u^2  - b^2 )^\alpha  }}{{u^2  - x^2 }}} f(u)du 
\quad \quad (x<b<e<f). 
\end{align}
Cooke also demonstrated the following valuable relations
\begin{align*}
&
\left( {\begin{array}{*{20}c}
   d  \\
   c  \\
\end{array}} \right)I_{\eta ,\alpha }^{ - 1} \left( {\begin{array}{*{20}c}
   x  \\
   c  \\
\end{array}} \right)I_{\eta ,\alpha }^{} f =  - \left( {\begin{array}{*{20}c}
   x & d  \\
   d & c  \\
\end{array}} \right)L_{\eta ,\alpha } f & (x > d > c)
\\ &
\left( {\begin{array}{*{20}c}
   e  \\
   d  \\
\end{array}} \right)K_{\eta ,\alpha }^{ - 1} \left( {\begin{array}{*{20}c}
   e  \\
   x  \\
\end{array}} \right)K_{\eta ,\alpha }^{} f =  - \left( {\begin{array}{*{20}c}
   d & e  \\
   x & d  \\
\end{array}} \right)M_{\eta ,\alpha } f & (x < d < e).
\end{align*}
These expressions are valid for $(-1<\alpha<1)$ and the most remarkable  outcome of their use is the disappearance of the troublesome integral derivatives.

It is useful to note that in many problems the solution implies solving an integral equation of Fredholm of second type whose kernel consists in the application of two double operators, for example
\begin{equation*}
\varphi (r) = \left( {\begin{array}{*{20}c}
   a & b  \\
   x & a  \\
\end{array}} \right)M_{\eta ,\alpha } \left( {\begin{array}{*{20}c}
   x & a  \\
   a & 0  \\
\end{array}} \right)L_{\bar \eta ,\bar \alpha } \varphi (x) + w(x) = \int_0^a {{\rm{K}}(x,u)\varphi (u)du + w(x)} \quad (0 < x < a).
\end{equation*}
Using the previous formulae, we shall have
\begin{align*}
{\rm{K}}(x,u) &= \frac{4}{{\pi ^2 }}\sin \pi \alpha \sin \pi \bar \alpha \frac{{(a^2  - u^2 )^{\bar \alpha } }}{{(a^2  - x^2 )^\alpha  }}x^{2(\eta  + \alpha )} u^{2\bar \eta  + 1} {\rm{H}}(x,u)
\\
{\rm{H}}(x,u) &= \int_a^b {\frac{{(t^2  - a^2 )^{\alpha  - \bar \alpha } }}{{(t^2  - x^2 )(t^2  - u^2 )}}} t^{1 - 2(\eta  + \bar \eta ) - 2\alpha } dt
\quad (0 < x < a)
\end{align*}
In many cases the last integral can be solved analytically.
\subsection{Combinations of Erd\'elyi-Kober and Hankel operators}\label{subsec:properties}
Based on the definitions of the operators, it is possible to show many important combination properties that are the decisive reason for reducing mixed boundary value problems to EK operators scheme. The reader can find the mathematical details and demonstrations in~\cite{J:1962:Sneddon62}. Here we limit ourselves to summarize them.
\subsubsection*{Combinations of operators of the same type}
\begin{align*}
I_{\eta ,\alpha } I_{\eta  + \alpha ,\beta }  &= I_{\eta ,\alpha  + \beta }  \\
K_{\eta ,\alpha } K_{\eta  + \alpha ,\beta }  &= K_{\eta ,\alpha  + \beta }  \\
S_{\eta  + \alpha ,\beta } S_{\eta ,\alpha }  &= I_{\eta ,\alpha  + \beta }  \\
S_{\eta ,\alpha } S_{\eta  + \alpha ,\beta }  &= K_{\eta ,\alpha  + \beta }  \\
S_{\eta ,0} S_{\eta ,0}  &= 1
\end{align*}
\subsubsection*{Combinations of operators of different type}
\begin{align*}
I_{\eta  + \alpha ,\beta } S_{\eta ,\alpha }  &= S_{\eta ,\alpha  + \beta }  \\
K_{\eta ,\alpha } S_{\eta  + \alpha ,\beta }  &= S_{\eta ,\alpha  + \beta }  \\
S_{\eta  + \alpha ,\beta } I_{\eta ,\alpha }  &= S_{\eta ,\alpha  + \beta }  \\
S_{\eta ,\alpha } K_{\eta  + \alpha ,\beta }  &= S_{\eta ,\alpha  + \beta }  
\end{align*}
In all cases the operators must have the same integration limits.
\subsection{Examples}
In tables~\ref{tab:A1}-\ref{tab:A2} the most useful operators used for solving the problems of this paper are made explicit for the default intervals $(0,x)$ and $(x,\infty)$. The necessary modifications for other intervals are evident.

\begin{table}[h!]
\renewcommand\arraystretch{2.5}
\caption{$\alpha=\frac{1}{2}$}
\begin{tabular}{|l|l|l|}
\hline
$\eta$  & $I_{\eta ,\frac{1}{2}}$ & $K_{\eta ,\frac{1}{2}}$ \\ [5pt] \hline
0  & $\frac{2}{{\sqrt \pi  }}\frac{1}{x}\int\limits_0^x {\frac{{t\,f(t)}}{{\sqrt {x^2  - t^2 } }}dt} $  & $\frac{2}{{\sqrt \pi  }}\int\limits_x^\infty  {\frac{{f(t)}}{{\sqrt {t^2  - x^2 } }}dt}$ \\ 
$-\frac{1}{4}$  & $\frac{2}{{\sqrt \pi  }}\frac{1}{{\sqrt x }}\int\limits_0^x {\frac{{t^{1/2} \,f(t)}}{{\sqrt {x^2  - t^2 } }}dt} $ & $\frac{2}{{\sqrt \pi  }}\frac{1}{{\sqrt x }}\int\limits_x^\infty  {\frac{{t^{1/2} f(t)}}{{\sqrt {t^2  - x^2 } }}dt} $ \\ 
$-\frac{1}{2}$  & $\frac{2}{{\sqrt \pi  }}\int\limits_0^x {\frac{{f(t)}}{{\sqrt {x^2  - t^2 } }}dt} $ & $
\frac{2}{{\sqrt \pi  }}\frac{1}{x}\int\limits_x^\infty  {\frac{{t\,f(t)}}{{\sqrt {t^2  - x^2 } }}dt}$ \\
\hline
\end{tabular}
\label{tab:A1}
\end{table}
\begin{table}[h!]
\renewcommand\arraystretch{2.5}
\caption{$\alpha=-\frac{1}{2}$}
\begin{tabular}{|l|l|l|}
\hline
$\eta$  & $I_{\eta ,-\frac{1}{2}}$ & $K_{\eta ,-\frac{1}{2}}$ \\ \hline 
0 & $\frac{1}{{\sqrt \pi  }}\frac{d}{{dx}}\int\limits_0^x {\frac{{t\,f(t)}}{{\sqrt {x^2  - t^2 } }}dt}$ & $
 - \frac{1}{{\sqrt \pi  }}\frac{1}{x}\frac{d}{{dx}}\int\limits_x^\infty  {\frac{{t^2 f(t)}}{{\sqrt {t^2  - x^2 } }}dt}$
\\ [5pt] 
$-\frac{1}{4}$ & $\frac{1}{{\sqrt \pi  }}\frac{1}{{\sqrt x }}\frac{d}{{dx}}\int\limits_0^x {\frac{{t^{3/2} \,f(t)}}{{\sqrt {x^2  - t^2 } }}dt} $ & $ - \frac{1}{{\sqrt \pi  }}\frac{1}{{\sqrt x }}\frac{d}{{dx}}\int\limits_x^\infty  {\frac{{t^{3/2} f(t)}}{{\sqrt {t^2  - x^2 } }}dt} $  \\
$-\frac{1}{2}$ & $\frac{1}{{\sqrt \pi  }}\frac{1}{x}\frac{d}{{dx}}\int\limits_0^x {\frac{{t^2 f(t)}}{{\sqrt {x^2  - t^2 } }}dt} $ & $ - \frac{1}{{\sqrt \pi  }}\frac{d}{{dx}}\int\limits_x^\infty  {\frac{{t\,f(t)}}{{\sqrt {t^2  - x^2 } }}dt}$ \\
\hline
\end{tabular}
\label{tab:A2}
\end{table}
%
\section{Magnetic field integral representation}\label{app_B}
In this appendix we derive in a simple way the Green functions necessary to express the magnetic field and the current density in integral form and then reformulate them by means of the Hankel operators $S_{\eta ,\alpha }$.

\subsection{Magnetic field generated by symmetric current lines}
Let us consider two infinite straight lines, parallel to the $z$-axis, and situated on the $x$-axis, in the position $x_0$ and $-x_0$, respectively, carrying the same current $J_0$. According to Biot-Savart law, the magnetic field they generate  is
\begin{align*}
H_x (x,y) = \frac{{J_0 }}{{2\pi }}\left( {\frac{{ - y}}{{(x - x_0 )^2  + y^2 }} + \frac{{ - y}}{{(x + x_0 )^2  + y^2 }}} \right) \\
H_y (x,y) = \frac{{J_0 }}{{2\pi }}\left( {\frac{{x - x_0 }}{{(x - x_0 )^2  + y^2 }} + \frac{{x + x_0 }}{{(x + x_0 )^2  + y^2 }}} \right).
\end{align*}
By using the Laplace transform these expressions can be rewritten in integral form. In the upper half space ($y>0$) we shall have
\begin{align*}
H_x (x,y) &=  - \frac{{J_0 }}{\pi }\int\limits_0^\infty  {e^{ - y\lambda } \cos (x\lambda )\cos (x_0 \lambda )d\lambda } \\
H_y (x,y) &= \frac{{J_0 }}{\pi }\int\limits_0^\infty  {e^{ - y\lambda } \sin (x\lambda )\cos (x_0 \lambda )d\lambda }. 
\end{align*}
In the case of two symmetric current line distributions ($J_z(-x)=J_z(x)$)  in the intervals $(-b,-a)$ and $a,b$ of the $x$-axis, we shall have by superposition 
\begin{align*}
H_x (x,y) =  - \frac{1}{\pi }\int\limits_0^\infty  {d\lambda \,e^{ - y\lambda } \cos (x\lambda )} \int\limits_a^b {J_z (\xi )\cos (\xi \lambda )d\xi }  \\
H_y (x,y) = \frac{1}{\pi }\int\limits_0^\infty  {d\lambda \,e^{ - y\lambda } \sin (x\lambda )} \int\limits_a^b {J_z (\xi )\cos (\xi \lambda )d\xi }. 
\end{align*}
Introducing the couple of cosine transforms
\begin{align*}
F(\lambda ) = \frac{1}{\pi }\int\limits_a^b {J_z (x)\cos (x\lambda )} dx
\quad  \leftrightarrow \quad
J_z (x) = 2\int\limits_0^\infty {F(\lambda )\cos (x\lambda )} d\lambda, 
\end{align*}
we can write for the upper half space ($y>0$)
\begin{align*}
H_x (x,y) &=  - \int\limits_0^\infty  {e^{ - y\lambda } \cos (x\lambda )F(\lambda )d\lambda } \\
H_y (x,y) &= \int\limits_0^\infty  {e^{ - y\lambda } \sin (x\lambda )} F(\lambda )d\lambda. 
\end{align*}
In the plane $y=0$ (where the current lines lie), on the upper face, we therefore shall have
\begin{align*}
H_x (x) &=  - \int\limits_0^\infty  {\cos (x\lambda )F(\lambda )d\lambda } \\
H_y (x) &= \int\limits_0^\infty  {\sin (x\lambda )} F(\lambda )d\lambda. 
\end{align*}
By means of the Hankel operators $S_{\eta ,\alpha }$, posing $F(\lambda ) = \lambda ^{1/2} \psi (\lambda )$, we can write
\begin{align*}
H_x (x) &=  - \left( {\pi x/2} \right)^{1/2} S_{ - 1/4,0} \psi (x) \\
H_y (x) &= (\pi x/2)^{1/2} S_{1/4,0} \psi (x) \\
J_z (x) &=  - 2H_x (x).
\end{align*}
Defining  $g(x) =  - \left( {\pi x/2} \right)^{ - 1/2} H_x (x)$ and $f(x) = \left( {\pi x/2} \right)^{ - 1/2} H_y (x)$, we shall have
\begin{align}
\label{eq:Hx_appB}
& S_{ - 1/4,0} \psi (x) = g(x) \\
\label{eq:Hy_appB}
& S_{1/4,0} \psi (x) = f(x) \\
\label{eq:Jz_appB}
& J_z (x) =  - 2H_x (x) = 2(\pi x/2)^{1/2} g(x).
\end{align}
According to the procedure explained in section~\ref{sec:dual_operator} this system is equivalent to the three modes listed in table~\ref{tab:tape_curr}.

\begin{table}[t!] 
\renewcommand\arraystretch{1.5}
\caption{Symmetric current distribution}
\label{tab:tape_curr}
\centering
\begin{tabular}{|l|l|l|} 
\hline
Mode I & Mode II & Mode III \\
\hline
$f = K_{  1/4,-1/2} \varphi $ & $f = I_{-1/4,  1/2} \varphi$ & $K_{-1/4, 1/2} f = I_{-1/4, 1/2} g$ \\ [5pt]
$g = I_{1/4,-1/2} \varphi $ & $g = K_{-1/4, 1/2} \varphi $ & $I_{1/4,-1/2} f = K_{1/4,-1/2} g$ \\ [5pt]
$\psi  = S_{1/4,-1/2} \varphi $ & $\psi  = S_{-1/4, 1/2} \varphi$  & \\
[5pt]
\hline
\end{tabular}
\end{table}

\subsection{Magnetic field generated by antisymmetric current lines}\label{sec:appB_antisymm}
In the case the current lines carry opposite currents $\pm J_0$, the magnetic field distribution given by Biot-Savart law is
\begin{align*}
H_x (x,y) &= \frac{{J_0 }}{{2\pi }}\left( {\frac{{ - y}}{{(x - x_0 )^2  + y^2 }} + \frac{y}{{(x + x_0 )^2  + y^2 }}} \right)
 \\
H_y (x,y) &= \frac{{J_0 }}{{2\pi }}\left( {\frac{{x - x_0 }}{{(x - x_0 )^2  + y^2 }} - \frac{{x + x_0 }}{{(x + x_0 )^2  + y^2 }}} \right).
\end{align*}
Following the same procedure detailed above, in the case of antisymmetric distribution   $J_z ( - x) =  - J_z (x)$ and using the sine transform couple, we have
\begin{align*}
F(\lambda ) = \frac{1}{\pi }\int\limits_a^b {J_z (x)\sin (x\lambda )} dx
\quad  \leftrightarrow \quad
J_z (x) = 2\int\limits_0^\infty {F(\lambda )\sin (x\lambda )} d\lambda
\end{align*}
and we can write for the upper half space ($y>0$)
\begin{align*}
H_x (x,y) &=  - \int\limits_0^\infty  {e^{ - y\lambda } \sin (x\lambda )F(\lambda )d\lambda } \\
H_y (x,y) &=  - \int\limits_0^\infty  {e^{ - y\lambda } \cos (x\lambda )} F(\lambda )d\lambda. 
\end{align*}
In the plane $y=0$ (where the current lines lie), on the upper face the field components become
\begin{align*}
H_x (x) &=  - \int\limits_0^\infty  {\sin (x\lambda )F(\lambda )d\lambda } 
\\
H_y (x) &=  - \int\limits_0^\infty  {\cos (x\lambda )} F(\lambda )d\lambda
\end{align*}
and again, by means of the Hankel operators  $S_{\eta ,\alpha }$, posing $F(\lambda ) = \lambda ^{1/2} \psi (\lambda )$, we can write
\begin{align*}
H_x (x) &=  - (\pi x/2)^{1/2} S_{1/4,0} \psi (x) \\
H_y (x) &=  - (\pi x/2)^{1/2} S_{ - 1/4,0} \psi (x).
\end{align*}
Defining $g(x) =  - \left( {\pi x/2} \right)^{ - 1/2} H_x (x)$ and $f(x) =  - \left( {\pi x/2} \right)^{ - 1/2} H_y (x)$, we shall have
\begin{align}
\label{eq:Hx_appB2}
& S_{1/4,0} \psi (x) = g(x) \\
\label{eq:Hy_appB2}
& S_{ - 1/4,0} \psi (x) = f(x) \\
\label{eq:Jz_appB2}
& J_z (x) =  - 2H_x (x) = 2(\pi x/2)^{1/2} g(x).
\end{align}
Again, according to the procedure explained in section~\ref{sec:dual_operator} this system is equivalent to the three modes listed in table~\ref{tab:tape_field}. 

\begin{table}[t!]
\renewcommand\arraystretch{1.5}
\caption{Antisymmetric current distribution}
\label{tab:tape_field}
\centering
\begin{tabular}{|l|l|l|}
\hline
Mode I & Mode II & Mode III \\
\hline
$f = K_{ - 1/4,1/2} \varphi $ & $f = I_{1/4, - 1/2} \varphi $ & $K_{1/4, - 1/2} f = I_{1/4, - 1/2} g$\\ [5pt]
$g = I_{ - 1/4,1/2} \varphi$ & $g = K_{1/4, - 1/2} \varphi $ & $I_{ - 1/4,1/2} f = K_{ - 1/4,1/2} g$\\ [5pt]
$\psi  = S_{-1/4, 1/2} \varphi$ & $\psi  = S_{1/4, -1/2} \varphi$ & \\
[5pt]
\hline
\end{tabular}
\end{table}
\subsection{Magnetic field generated by circular currents}\label{appB_circ}
Let us consider a coil of radius $a$, centered in the origin of a system of cylindrical coordinates, carrying a current $I_0$. It is well know that it generates a magnetic field that can be derived from a vector potential ${\bf{A}} = (0,A_\varphi  ,0)$, with
\begin{align*}
A_\varphi  (r,z) = \frac{{I_0 }}{\pi }\sqrt {\frac{a}{r}} \frac{1}{k}\left[ {\left( {1 - \frac{{k^2 }}{2}} \right){\rm K}(k) - {\rm E}(k)} \right]\quad \left( {k^2  = \frac{{4ar}}{{(a + r)^2  + z^2 }}} \right),
\end{align*}
where ${\rm K}(k)$ and ${\rm E}(k)$ are the complete elliptic integrals of the first and second kind. By derivation we obtain the components of the magnetic field
\begin{align*}
& H_r (r,z) =  - \frac{{\partial A_\varphi  }}{{\partial z}} = \frac{{I_0 }}{{2\pi }}\frac{1}{{\sqrt {(a + r)^2  + z^2 } }}\frac{z}{r}\left[ { - {\rm K}(k) + \frac{{a^2  + r^2  + z^2 }}{{(a - r)^2  + z^2 }}{\rm E}(k)} \right]\\
& H_z (r,z) = \frac{1}{r}\frac{{\partial (rA_\varphi  )}}{{\partial r}} = \frac{{I_0 }}{{2\pi }}\frac{1}{{\sqrt {(a + r)^2  + z^2 } }}\left[ {{\rm K}(k) + \frac{{a^2  - r^2  - z^2 }}{{(a - r)^2  + z^2 }}{\rm E}(k)} \right].
\end{align*}
By using the Hankel transform, it is possible to rewrite these expressions in an integral form. For $z>0$ one has
\begin{align*}
& A_\varphi  (r,z) = \frac{{\mu I_0 a}}{2}\int\limits_0^\infty  {e^{ - z\lambda } J_1 (r\lambda )} J_1 (a\lambda )d\lambda  \\
& H_r (r,z) = \frac{{I_0 a}}{2}\int\limits_0^\infty  {e^{ - z\lambda } J_1 (r\lambda )} J_1 (a\lambda )\lambda d\lambda \\
& H_z (r,z) = \frac{{I_0 a}}{2}\int\limits_0^\infty  {e^{ - z\lambda } J_0 (r\lambda )} J_1 (a\lambda )\lambda d\lambda. 
\end{align*}
In the case of an azimuthal current distribution $J_\phi  (r)$ flowing in a thin disc of radius $a$ and situated in the plane $z=0$, we shall have by superposition
\begin{align*}
& H_r (r,z) = \frac{1}{2}\int\limits_0^\infty  {J_1 (r\lambda )e^{ - z\lambda } \lambda d\lambda } \int\limits_0^a {J_\phi  (\rho )J_1 (\rho \lambda )\rho d\rho } 
\\
& H_z (r,z) = \frac{1}{2}\int\limits_0^\infty  {J_0 (r\lambda )e^{ - z\lambda } \lambda d\lambda } \int\limits_0^a {J_\phi  (\rho )J_1 (\rho \lambda )\rho d\rho }. 
\end{align*} 
Defining the couple of Hankel transforms
\begin{align*}
\psi(\lambda ) = \frac{1}{2}\int\limits_0^a {J_\phi  (\rho )} J_1 (\lambda \rho )\rho d\rho 
\quad \leftrightarrow \quad
J_\phi  (\rho ) = 2\int\limits_0^\infty {\psi(\lambda )} J_1 (\lambda \rho )\lambda d\lambda 
\end{align*}
we can also write
\begin{align*}
& H_r (r,z) = \int\limits_0^\infty  {F(\lambda )J_1 (r\lambda )e^{ - z\lambda } \lambda d\lambda } 
\\
& H_z (r,z) = \int\limits_0^\infty  {F(\lambda )J_0 (r\lambda )e^{ - z\lambda } \lambda d\lambda }.
\end{align*}
In the plane of the disc ($z=0$), on the upper face, we shall therefore have
\begin{align*}
H_r (r) &= \int\limits_0^\infty  {F(\lambda )J_1 (r\lambda )\lambda d\lambda } \\
H_z (r) &= \int\limits_0^\infty  {F(\lambda )J_0 (r\lambda )\lambda d\lambda }. 
\end{align*}
Again, by means of Hankel operators $S_{\eta ,\alpha }$, posing $F(\lambda ) = \lambda ^{1/2} \psi (\lambda )$, we can write
\begin{align}
& S_{1/2,0} \psi (r) = H_r (r) = g(r) \label{eq:appB_circ1}
\\
& S_{0,0} \psi (r) = H_z (r) = f(r) \label{eq:appB_circ2}
\\
& J_\phi  (r) = 2H_r (r) = 2g(r).
\end{align}
Again, according to the procedure explained in section~\ref{sec:dual_operator} this system is equivalent to the three modes listed in table~\ref{tab:disc_unif}.
\begin{table}[t!] 
\renewcommand\arraystretch{1.5}
\caption{Circular current distribution}
\label{tab:disc_unif}
\centering
\begin{tabular}{|l|l|l|}
\hline
Mode I & Mode II & Mode III \\ 
\hline
$f = K_{0,1/2} \varphi $ & $f = I_{1/2, - 1/2} \varphi$ & $K_{1/2, - 1/2} f = I_{1/2, - 1/2} g $ \\ [5pt]
$g = I_{0,1/2} \varphi$ & $g = K_{1/2, - 1/2} \varphi $ & $I_{0,1/2} f = K_{0,1/2} g $ \\ [5pt]
$\psi  = S_{0,1/2} \varphi$ & $\psi = S_{1/2, - 1/2} \varphi $ & \\ [5pt]
\hline
\end{tabular}
\label{tab:disc}
\end{table}

\end{document}